\documentclass[preprint2]{aastex}

\newcommand{\be}{\begin{equation}}
\newcommand{\ee}{\end{equation}}
\newcommand{\ben}{\begin{eqnarray}}
\newcommand{\een}{\end{eqnarray}}
\newcommand{\bc}{\begin{center}}
\newcommand{\ec}{\end{center}}

\shorttitle{A supermassive boson star at the galactic center?}
\shortauthors{Torres, Capozziello, Lambiase}

\begin{document}

\title{A supermassive scalar star at the Galactic Center?}


\author{Diego F. Torres}
\affil{Instituto Argentino de Radioastronom\'{\i}a, C.C.5, 1894
Villa Elisa, Buenos Aires, Argentina}
\email{dtorres@venus.fisica.unlp.edu.ar}

\and

\author{S. Capozziello and G. Lambiase}
\affil{Dipartimento di Scienze Fisiche E.R. Caianiello,
Universit\`a di Salerno, 84081 Baronissi (SA), Italy\\ Istituto
Nazionale di Fisica Nucleare, Sez. Napoli,
Italy}\email{capozziello@sa.infn.it}\email{lambiase@sa.infn.it}

\begin{abstract}
We explore whether supermassive non-baryonic stars (in particular
boson, mini-boson and non-topological soliton stars) might be at
the center of some galaxies, with special attention to the Milky
Way. We analyze, from a dynamical point of view, what current
observational data show, concluding that they are compatible with
a single supermassive object without requiring it to be a black
hole. Particularly, we show that scalar stars fit very well into
these dynamical requirements. The parameters of different models
of scalar stars necessary to reproduce the inferred central mass
are derived, and the possible existence of boson particles with
the adequate range of masses is commented. Accretion to boson
stars is also analyzed, and a comparison with another non-baryonic
candidate, a massive neutrino ball, which is also claimed as an
alternative to the central black hole, is given. Both models are
capable to explain the nature of the object in Sgr A$^*$ without
invoking the presence of a singularity. One difficult issue is why
the accreted materials will not finally produce, in a sufficiently
long time, a black hole. We provide an answer based on stellar
disruption in the case of boson stars, and comment several
suggestions for its possible solution in neutrino ball scenarios.
Finally, we discuss the prospects for the observational detection
of these supermassive scalar objects, using the new generation of
X-ray and radio interferometry satellites.

\end{abstract}


\newpage

\section{Introduction}

During the last years, the possible existence of a single large
mass in the Galactic Center has been favored as the upper bound on
its size tightens, and stability criteria rule out complex
clusters. Although it is commonly believed that this central mass
is a supermassive black hole, it is not yet established, as we
discuss below, on a firm observational basis.

The aim of this paper is to present an alternative model for the
supermassive dark object in the center of our Galaxy, formed by
self-gravitating non-baryonic matter composed by bosons. This kind
of objects, so-called boson stars, are well known to physicists,
but up to now, observational astrophysical consequences have
hardly been explored. The main characteristics of this model are:
\begin{enumerate}
  \item It is highly relativistic, with a size comparable to (but
  slightly larger than) the Schwarzschild radius of a black hole
  of equal mass.
  \item It has neither an event horizon nor a singularity,
  and after a physical radius is reached, the
  mass distribution exponentially decreases.
  \item The particles that form the object interact between each other
  only gravitationally, in such a way that there is no solid
  surface to which falling particles can collide.
\end{enumerate}

It is the purpose of this work to show that these features are
able to produce a Galactic Center model which can be confronted
with all known observational constraints, and to point the way in
which such a center could be differentiated from a usual
supermassive black hole.

\subsection{Why scalar fields?}

Interesting models for dark matter use weakly interacting bosons,
and primordial nucleosynthesis show that most of the mass in the
universe should be non-baryonic if $\Omega \sim 1$. Most models of
inflation make use of scalar fields. Scalar-tensor gravitation is
the most interesting alternative to general relativity. Recent
results from supernovae, which in principle were thought of to
favor a cosmological constant \citep{per99}, can as well be
supported by a variety of models, some of them with scalar fields
too \citep{cel99}. Particle physicist expect to detect the scalar
Higgs particle in the next generation of accelerators. Scalar
dilatons appear in low energy unified theories, where the tensor
field $g_{\mu\nu}$  of gravity is accompanied by one or several
scalar fields, and in string effective super-gravity. The axion is
a scalar with a long history as a dark matter candidate, and
Goldstone bosons have also already inferred masses. Symmetry
arguments, which once led to the concept of neutron stars, may
force to ask whether there could be stellar structures made up of
bosons instead of fermions.

In recent works, Schunck and Liddle (1997), Schunck and Torres
(2000), and Capozziello, Lambiase and Torres (2000) analyzed some
of the observational properties of boson stars, and found them
notoriously similar to black holes. In particular, the
gravitational redshift of the radiation emitted within a boson
star potential, and the rotational curves of accreted particles,
were studied to assess possible boson star detection.  The
interest in observational properties of boson stars also led to
investigate them as a possible lens in a gravitational lensing
configuration \citep{dab00}. Recent studies are putting forward
the gravitational lensing phenomenon in strong field regimes
\citep{vir98,vir00,tor98a,tor98b}. This would be the case for
boson stars, which are genuine relativistic objects.

This work is organized as follows. Sec. II is a brief summary of
the main observational results concerning Sgr A$^*$, and the main
hypotheses in order to explain it are outlined there. Sec. III
analyzes what dynamical observational data show, and what kind of
models can support them. Sec. IV gives the basic ingredients to
theoretically construct scalar stars, shows mass and radius
estimates, and study effective potentials and orbits of particles.
In Sec. V we study the center of the Galaxy, show which are the
stable scalar star models able to fit such a huge mass, and
comment on the possible existence of boson particles with the
required features. We also provide there an assessment of the
disruption processes in boson star scenarios. Discussion and
conclusions are given in Sections VI and VII.

\section{The Galactic Center}

\subsection{Main observational facts}

The Galactic Center is a very active region toward the Sagittarius
constellation where, at least, six very energetic radio sources
are present (Sgr A, A$^*$, B1, B2, C, and D). Furthermore, there
are several supernova remnants, filaments, and very reach star
clusters.  Several observational campaigns \citep{gen94} have
identified the exact center with the supermassive compact dark
object in Sgr A$^*$, an extremely loud radio source. The mass and
the size of the object has been established to be $(2.61\pm
0.76)\times 10^6 M_{\odot}$ concentrated within a radius of 0.016
pc (about 30 lds)\citep{ghe98,gen96}.

More precisely, Ghez et al. (1998) have made extremely accurate
velocity measurements in the central square arcsecs. From this
bulk of data, it is possible to state that a supermassive compact
dark object is present at the Galactic Center. It is revealed by
the motion of stars moving within projected distances around 0.01
pc from the radio source Sgr A$^*$, at projected velocities in
excess of 1000 km s$^{-1}$.  In other words, a high increase in
the velocity dispersion of the stars toward the dynamical center
is revealed. Furthermore, a large and coherent counter--rotation,
especially of the early--type stars, is shown, supporting their
origin in a well--defined epoch of star formation. Observations of
stellar winds nearby Sgr A$^*$ give a mass accretion rate of $
dM/dt=6\times 10^{-6}M_{\odot}$ yr$^{-1}$ \citep{gen96}. Hence,
the dark mass must have a density $\sim 10^9 M_{\odot}$ pc$^{-3}$
or greater, and a mass--to--luminosity ratio of at least
$100M_{\odot}/L_{\odot}$. The bottom line is that the central dark
mass seems to be a single object, and that it is statistically
very significant $(\sim 6-8\sigma)$.

\subsection{Domain of the black hole?}

Such a large density contrast excludes that the dark mass could be
a cluster of almost $2\times 10^6$ neutron stars or white dwarfs.
Detailed calculations of evaporation and collision mechanisms give
maximal lifetimes of the order of $10^8$ years, much shorter than
the estimated age of the Galaxy \citep{san92,mao95}.

As a first conclusion, several authors state that in the Galactic
Center there is either a single supermassive black hole or a very
compact cluster of stellar size black holes \citep{gen96}. We
shall come back to this paradigm through the rest of the paper
(particularly in Section III).

Dynamical evidence for central dark objects has been published for
17 galaxies, like M87 \citep{for94,mac97}, or NGC4258
\citep{gre95}, but proofs that they really are black holes
requires measurement of relativistic velocities near the
Schwarzschild radius, $r_s \simeq 2 M_\bullet/(10^8~M_\odot)$ AU
\citep{kor95}, or seeing the properties of the accretion disk. Due
to the above mentioned mass accretion rate, if Sgr A$^*$ is a
supermassive black hole, its luminosity should be more than
$10^{40}$erg s$^{-1}$, provided the radiative efficiency is about
10\%. On the contrary, observations give a bolometric luminosity
less than $10^{37}$ erg s$^{-1}$, already taking into account the
luminosity extinction due to interstellar gas and dust. This
discrepancy is the so--called ``blackness problem'' which has led
to the notion of a ``black hole on starvation''. Standard dynamics
and thermodynamics of the spherical accretion onto a black hole
must be modified in order to obtain successful models
\citep{fal94}. Recent observations, we recall, probe the
gravitational potential at a radius larger than $4\times 10^{4}$
Schwarzschild radii of a black hole of mass $2.6\times
10^{6}M_{\odot}$ \citep{ghe98}.

\subsection{Domain of the massive neutrino?}

An alternative model for the supermassive compact object in the
center of our Galaxy has been recently proposed by Tsiklauri and
Viollier (1998). The main ingredient of the proposal is that the
dark matter at the center of the galaxy is non-baryonic (but in
any case fermions, e.g. massive neutrinos or gravitinos),
interacting gravitationally to form supermassive balls in which
the degeneracy pressure balances their self--gravity. Such
neutrino balls could have formed in early epochs, during a
first--order gravitational phase transition and their dynamics
could be reconciled, with some adjustments, to the Standard Model
of Cosmology.

Several experiments are today running to search for neutrino
oscillations. LSND \citep{whi98} found evidence for oscillations
in the $\nu_{e}-\nu_{\mu}$ channel for pion decay at rest and in
flight. On the contrary KARMEN \citep{zei98} seems to be in
contradiction with LSND evidence. CHORUS and NOMAD at CERN are
just finishing the phase  of 1994--1995 data analysis. It is very
likely that exact predictions for and $\nu_{\mu}-\nu_{\tau}$
oscillations will be available in next years. Thanks to the fact
that it is possible to give correct values for the masses to the
quarks {\it up}, {\it charm}, and {\it top}, it is possible to
infer reasonable values of mass for $\nu_{e}$, $\nu_{\mu}$, and
$\nu_{\tau}$. In order to explain the characteristics of Sgr
A$^*$, we need fermions whose masses range between 10 and 25
keV which, cosmologically, fall into the category of warm dark
matter. It is interesting to note that a good estimated value for
the mass of $\tau$-neutrino is
\be
m_{{\nu_{\tau}}}=m_{{\nu_{\mu}}}\left(\frac{m_{t}}{m_{c}}\right)^{2}
\simeq 14.4\;\; {\rm keV}\,. \ee Choosing
fermions like neutrinos or gravitinos in this mass
range allows for the formation of supermassive degenerate objects
from $10^6 M_{\odot}$ to $10^9 M_{\odot}$.

The theory of heavy neutrino condensates, bound by gravity, can be
easily sketched considering a Thomas--Fermi model for fermions
\citep{vio92}.  We can set the Fermi energy
 equal to the gravitational potential which binds the
system, and  see that the number density is a function of the
gravitational potential. Such a gravitational potential will obey
a Poisson equation where neutrinos (and anti-neutrinos) are the
source term. This equation is valid everywhere except at the
origin. By a little algebra, the Poisson equation reduces itself
to the radial Lan\'e--Emden differential equation with polytropic
index $n=3/2$, which is equivalent to the Thomas--Fermi
differential equation of atomic physics, except for a minus sign
that is due to the gravitational attraction of the neutrinos. The
Newtonian potential, which is the solution of such an equation, is
$\Phi(r)\sim r^{-4}$. Considering a standard accretion disk, if
Sgr A$^*$ is a neutrino star with radius $R=30.3$ lds ($\sim 10^5$
Schwarzschild radii), mass $M_{GC}=2.6\times 10^{6}M_{\odot}$, and
luminosity $L_{GC}\sim 10^{37}$erg sec$^{-1}$, it should consist
of neutrinos with masses $m\geq 12.0$ keV for $g_{\nu}=4$, or
$m\geq 14.3$ keV for $g_{\nu}=2$ \citep{bil98,vio92}. It is
specially appealing that with the same mass for neutrinos, several
galactic centers can be modeled. For instance, similar results
hold also for the dark object ($M\sim 3\times 10^{9}M_{\odot} $)
inside the center of M87. Due to the Thomas--Fermi theory, the
model fails at the origin: we have to consider the effect of the
surrounding baryonic matter which, in some sense, has to stabilize
the neutrino condensate. In fact, the solution $\Phi(r)\sim
r^{-4}$ is clearly unbounded from below.

The model proceeds assuming a thermodynamical phase where a
constant neutrino number density can be taken into consideration.
This is quite natural for a Fermi gas at temperature $T=0$. The
Poisson equation can be recast in a Lan\'e--Emden form with
polytropic index $n=0$. The solution of such an equation is
$\Phi(r)\sim r^{2},$ which is clearly bounded \citep{cap99}. For
$T\neq 0$,  we get a solution of the form $\Phi(r)\sim r^{-4}$.
Matching these two results it is possible to confine the neutrino
ball. On the other hand, a similar result is recovered using the
Newton Theorem for a spherically symmetric distribution of matter
of radius $R$ \citep{bin87}. In that case, the potential goes
quadratic inside the sphere while it goes as $\Phi(r)\sim r^{-1}$
matching on the boundary. In our case, the situation is similar
assuming the matching with a steeper potential.

If such a neutrino condensate exists in the center of Galaxy, it
could act as a spherical thick lens (a magnifying glass) for the
stars behind it, so that their apparent velocities will be larger
than in reality. In other words, depending on the line of sight,
it should be possible to correct the projected velocities by a
gravitational lensing contribution, so trying to explain the
bimodal distribution (early and late type stars) actually observed
\citep{ghe98,gen96}. Since the astrophysical features of the
object in Sgr A$^*$ are quite well known, accurate observations by
lensing could contribute to the exact determination of particle
constituents. A detailed model and comparison with the data was
presented by some of us \citep{cap99}.

\section{Observational status}

\subsection{Brief review on dynamical data and what are they
implying}

An important information on the central objects of galaxies
(particularly in active galactic nuclei) is the short timescale of
variability. This has the significance of putting an upper bound
--known as causality constraint-- on the size of the emitting
region: If a system of size $L$ suddenly increases its emissivity
at all points, the temporal width with which we receive it is
$L/c$ and thus, a source can not fluctuate in a way that involves
its entire volume in timescales shorter than this (unless $c$ is
not a limiting velocity). This finally yields a corresponding
maximum lenghscale, typically between $10^{-4}$ to 10 pc
\citep{kro99}. Autocorrelation in the emission argue against the
existence of a cluster of objects (unless only one member
dominates the emissivity), and the small region in which the
cluster should be allows two body encounters to be very common and
produces the lost of the stability of complex clusters. These
facts can be used to conclude that a single massive object must be
in the center of most galaxies. {\it What observations show in
this case is that the size of the emitting regions are very
small}. This does not directly implicate black holes as such.

The way in which we expect to detect the influence of a very
massive object is through its gravity. The ``sphere of influence''
of a large mass is defined by the distance at which its potential
significantly affects the orbital motions of stars and gas, and is
given by \be R_*=GM/\sigma_*^2 \sim 4 M_7 \sigma_{*,100}^{-2} {\rm
pc}, \ee where the central mass is normalized to $10^{7} M_\odot$
and $\sigma_{*,100}$ is a rms orbital speed in units of 100 km
s$^{-1}$. Thus, even very large masses have a small sphere of
influence. Within this sphere, the expected response to a large
nuclear mass can be divided in two groups. Firstly, the response
of interstellar gas can hardly be anything different from an
isotropic motion in the center of mass frame. Random
speeds are often thought to be much smaller than this overall
speed, which far from the sphere of influence is just given by
$(GM/r)^{1/2}$.  If the energy produced by random motions is
radiated away, the gas behaves as a whole and tend to flatten
itself, conserving the angular momentum. Observations of the
rotational velocity of gas as a function of the radius can thus
provide a measure of the total mass at the center. Then, {\it what
observation searches is the existence of a Keplerian potential
signature in the flattened gas velocity distribution.} Examples of
this are the radio galaxies M87 \citep{for94}, NGC 4258
\citep{miy95}, and many others. Then, any massive object
producing a velocity distribution with a Keplerian decrease would
be allowed by observation.

Secondly, we consider the response of the stars. In this case,
peculiar velocities are larger or comparable to the bulk streaming
and it is harder to actually differentiate the stellar response to
a nuclear mass from the observable properties of a pure stellar
potential. However, stellar motions, contrary to that of gas, is
not affected by other forces (as those produced by magnetic
fields), and are better tracers of mass distributions. {\it What
observations show in this case are the stellar density and the
velocity distribution}. Use of the collisionless Boltzmann
equation allows one to get Jean's relation \citep{bin87,kro99}:
\be \label{JEANS}\frac{GM(r)}{r}= V_{rot}^2-\sigma_r^2 \left[
\frac{d \ln n(r)}{d \ln r} - \frac{d \ln \sigma_r^2}{d \ln r} + 2
- \frac{\sigma_t^2}{ \sigma_r^2} \right] \ee here, $n(r$) is the
spherically symmetric distribution of stars, $V_{rot}$ is the mean
rotational speed, $\sigma_{r,\theta,\phi}$ are velocity
dispersions and $\sigma_t^2=\sigma_\theta^2+\sigma_\phi^2$.
Measurements of $V$ and random velocities determine that $M(r)$
decline inwards until a critical radius, where $M$ becomes
constant. What really happens, however, if that even if this
region is not scrutinized, the mass to light ratio becomes
sufficiently large to suggest the presence of a large dark mass. A
combination of this with other measurements {\it strongly suggest
that this dark mass is a single object.}

To summarize, usual measurements point to inform us that there
exist a single supermassive object in the center of some galaxies
but do not definitively state its nature. As Kormendy and
Richstone (1995) have stated, the black hole scenario has become
our paradigm. But suggestions that the dark objects are black
holes are based only on indirect astrophysical arguments, and
surprises are possible on the way to the center.

\subsection{The Galaxy}

We follow Genzel et al. (1996) and parameterize the stellar
density distribution as, \be \label{eq0} n(r) = \frac
{\Sigma_0}{R_0} \frac 1{1+ (R/R_0)^\alpha}. \ee Note that $R_0$ is
related to the core radius through $R_{{\rm core}}=b(\alpha) R_0$.
Genzel et al. found that the best fit parameters for the observed
stellar cluster are a central density of $4\times 10^6
M_\odot$pc$^{-3}$, a core radius of 0.38 pc, and a value
$\alpha=1.8$ ($b(\alpha)=2.19$). With this distribution, they
found that a dark mass of about 2.5 $\times 10^{6} M_\odot$ was
needed to fit the observational data.

The cluster distribution, and the cluster plus a black hole
constant mass is shown in Fig. \ref{fiteo}. Black boxes represent
Genzel et al. 1996 (Table 10) and Eckart and Genzel 1997 (Fig. 5)
data. A dashed line stands for the stellar cluster contribution,
while a dot-dashed line represent an enclosed point-like black
hole mass. The solid line in the right half of the figure stands
for the mass distribution both for a black hole and a boson star
plus the stellar cluster. The mass dependence we are plotting for
the boson star is obtained in Section IV, and represents the mass
distribution of a mini-boson star ($\Lambda=0$, $m[{\rm GeV}]=2.81
\times 10^{-26}$) with dimensionless central density $\sigma(0)$
equal to 0.1. Other boson star configurations, with appropriate
choice of the boson field mass $m$, yield to the same results.
Note the break of more than three orders of magnitude in the
x-axis, this is caused because the boson star distribution,
further out of the equivalent Schwarzschild radius, behaves as a
black hole. It begins to differ from the black hole case at radius
more than three orders of magnitude less than the innermost data
point, that is why the x-axis has to have a break. From the mass
distribution, a boson star in the center of the galaxy is
virtually indistinguishable from a black hole.

Tsiklauri and Viollier (1998) have shown that the same
observational data can also be fitted using an extended neutrino
ball. In that case, differences begin to be noticed just around
the innermost data point. It is then hard to determine whether the
central object is a black hole, a neutrino ball, or a boson star
based {\it only} on dynamical data now at hand, and other problems
concerning the accretion disk have to considered (see below). Even
harder is the situation for deciding --using only this kind of
data-- if the supermassive object is a boson star instead of a
black hole: as a boson star is a relativistic object, the decay of
the enclosed mass curve happens close the center. This, however,
will provide an equivalent picture than a black hole for
disruption and accretion processes; we shall comment on it in the
next sections.

To apply Eq. (\ref{JEANS}) to the observational data we have to
convert intrinsic velocity dispersions ($\sigma_r(r)$) and volume
densities ($n(r)$) to projected ones \citep{bin87}, these are the
ones we observe. We shall also consider that $\delta = 1
-\sigma_\theta^2 / \sigma_r^2$, the anisotropy parameter, is equal
to 0, and we are assuming implicitly that $\sigma_\theta =
\sigma_\phi$. We take into account the following Abel integrals,
\be \label{eq1} \Sigma(p)= 2 \int _p ^\infty \frac{ n(r) r}{
\sqrt{r^2 -p^2}} dr , \ee \be  \label{eq2}
 \Sigma(p)\sigma_r(p)= 2 \int _p ^\infty \frac{ n(r)
\sigma_r(r)^2 r}{ \sqrt{r^2 -p^2} } dr . \ee $\Sigma(p)$ denotes
surface density, $\sigma_r(p)$ is the projected velocity
dispersion, and $p$ is the projected distance. We adopt Genzel et
al.'s (1996) and Tsiklauri and Viollier's (1998) parameterization
for $\sigma_r(r)$, \be \sigma_r(r)=\sigma(\infty)^2 +
\sigma(2'')^2 \left(\frac{R}{2''}\right)^{-2\beta}.\ee What one
usually does is to numerically integrate Eqs.
(\ref{eq1},\ref{eq2}) and fit the observational data ($\sigma_r(p)
\;\;vs. \;\;p$). To do so one also has to assume a density
dependence for the cluster (as in Eq. (\ref{eq0})), and the dark
mass. With a point-like dark mass of 2.5 $\times 10^6 M_\odot$,
the parameters of the fit result to be $\sigma(\infty)=55$ km
s$^{-1}$, $\sigma(2'')=350$ km s$^{-1}$, and $\beta = 0.95$.

If one now changes the central black hole for a neutrino ball, or
a boson star, one has to consider the particular density
dependence for these objects. In this way, $n(r)=n_{{\rm
stellar\;cluster}}(r) + n_{{\rm dark \;mass} }(r)$. We obtained
$n_{{\rm boson \;star }}(r)$ as $M(r)/(4/3 \pi r^3)$, where $M(r)$
is the fitted mass dependence of the boson star as explained in
the next section. It is now useful to consider that the fitting of
$\sigma_r(p)$ is made taking into account observational data
points in regions where the boson star generated space-time is
practically indistinguishable from a black hole. Then, we may
expect that the actual parameters, $\sigma(\infty)$,
$\sigma(2'')$, and $\beta$, will be very close to those obtained
for the black hole. For our purposes, it is enough to take the
same $\sigma_r(r)$ as in the black hole case, and compute
$\sigma_{r\;\;{\rm boson \;star}}(p)$ using Eqs.
(\ref{eq1},\ref{eq2}), with the adequate total $n(r)$.

In Fig. \ref{2deTSI} we show the observational data of Eckart and
Genzel (1997) (filled black boxes) and Genzel et al. (1996)
(hollow circles) super-imposed with the curve for $\sigma_r(p)$
that we obtain with a mini-boson star ($\Lambda=0$,
$\sigma(0)=0.1$, $m[{\rm GeV}]=2.81 \times 10^{-26}$) in the
galactic center. Other boson star configurations, with appropriate
choice of the boson field mass $m$, yield to the same results. For
this configuration, we used, as will be explained in Section IV, a
mass distribution given by a Boltzmann-like equation with $A_2=2.5
\times 10^6 M_\odot$, $A_1=-0.237 \times 10^6 M_\odot$, $R_0=1.192
10^{-6}$ pc, and $\Delta R=4.163 10^{-6}$ pc. In the range
plotted, and in which data is available, the differences between
boson and black hole theoretical curves is undetectable.  They
only begin to deviate from each other at $p\sim 10^{-4}$ pc, well
beyond the last observational data point. Even the deviation in
such a region is as slight as 1 km s$^{-1}$, and it only becomes
more pronounced when $p$ values are closer to the center. However,
we should recall that it has no sense to go to such extreme values
of $p$: the stars will be disrupted by tidal forces (see the
discussion in Section VI) in those regions.

\section{Boson stars}

\subsection{Basic concepts and configurations}

Let us study the Lagrangian density of a massive complex
self-gravitating scalar field (taking $\hbar=c=1$),

\be
{\cal L} = \frac {1}{2} \sqrt{\mid g \mid} \left[
  \frac {m_{{\rm Pl}}^2}{8\pi} R + \partial_\mu \psi^\ast \partial^\mu \psi
- U(|\psi |^2) \right ] \; , \label{lagr} \ee where $R$ is the
scalar of curvature, $|g|$ the modulus of the determinant of the
metric $g_{\mu \nu}$, and $\psi$ is a complex scalar field with
potential $U$. Using this Lagrangian as the matter sector of the
theory, we get the standard field equations, \ben R_{\mu \nu } -
\frac{1}{2} g_{\mu \nu } R & = &
                  - \frac{8\pi}{m_{{\rm Pl}}^2} T_{\mu \nu } (\psi ) \; , \\
\Box \psi + \frac{dU}{d|\psi |^2} \psi & = & 0 \; ,
\een
where the stress energy tensor is given by,
\begin{eqnarray}
T_{\mu \nu } & = & (\partial_\mu \psi^\ast ) (\partial_\nu \psi )+
    \nonumber \\
& &  - \frac{1}{2} g_{\mu \nu }
 \Bigl [ g^{\alpha \beta } (\partial_\alpha \psi^\ast )
         (\partial_\beta \psi ) - U(|\psi |^2) \Bigr ]
\end{eqnarray}
and
\be
\Box = \partial_\mu \Bigl [ \sqrt{\mid g \mid } \; g^{\mu \nu }
\partial_\nu \Bigr ]/ \sqrt{\mid g \mid } \ee is the covariant
d'Alembertian. Because of the fact that the potential is a
function of the square of the modulus of the field, we obtain a
global $U(1)$ symmetry. This symmetry, as we shall later discuss,
is related with the conserved number of particles. The particular
form of the potential is what makes the difference between
mini-boson, boson, and soliton stars. Conventionally, when the
potential is given by
\be
U = m^2 |\psi |^2 + \frac{\lambda }{2} |\psi |^4 \; , \ee where
$m$ is the scalar mass and $\lambda$ a dimensionless constant
measuring the self-interaction strength, mini-boson stars are
those spherically symmetric equilibrium configurations with
$\lambda =0 $. Boson stars, on the contrary, have a non-null value
of $\lambda$. The previous potential with $\lambda \neq 0$ was
introduced by Colpi et al. (1986), who numerically found that the
masses and radius of the configurations were deeply enlarged in
comparison to the mini-boson case.

Soliton (also called non-topological soliton) stars are different
in the sense that, apart from the requirement that the Lagrangian
must be invariant under a global $U(1)$ transformation, it is
required that --in the absence of gravity-- the theory must have
non-topological solutions; i.e. solutions with a finite mass,
confined to a finite region of space, and non-dispersive. An
example of this kind of potentials is the one introduced by Lee
and his coworkers \citep{lee87},
\be
U=m^2 |\psi|^2 \left(1 - \frac{|\psi|^2}{\Phi_0^2}\right)^2, \ee
where $\Phi_0$ is a constant. In general, boson stars accomplish
the requirement of invariance under a $U(1)$ global transformation
but not the {\it solitonic} second requirement. To fulfill it, it
is necessary that the potential contains attractive terms. This is
why the coefficient of $(\psi^* \psi)^2$ of the Lee form has a
negative sign. Finally, when $|\psi| \rightarrow \infty$, $U$ must
be positive, which leads, minimally, to a sixth order function of
$\psi$ for the self-interaction.  It is usually assumed, because
of the range of masses and radius for soliton stars in
equilibrium, that they are huge and heavy objects, although this
finally depends on the choice of the different parameters.

We shall now briefly explain how these configurations can be
obtained \citep{kau68,ruf69,lee92,lid92,mie98}. We adopt a
spherically symmetric line element
\be
ds^2 = e^{\nu (r)} dt^2 - e^{\mu (r)} dr^2
  - r^2 ( d\vartheta^2 + \sin^2\vartheta \, d\varphi^2) \;,
\ee
with a scalar field time-dependence ansatz consistent with this metric:
\be
\psi (r,t) = \sigma(r) e^{-i \omega t} \; \ee where $\omega$ is
the (eigen-)frequency. This form of the field ensures us to be
working in the configurations of minimal energy \citep{lee87}.

The non-vanishing components of the energy-momentum tensor are
\ben T_0{}^0 = \rho = \frac{1}{2} [ \omega^2  \sigma^2(r) e^{-\nu
}
   + \sigma'^2(r) e^{-\mu } + U ] \; , \\
T_1{}^1 = p_r = \frac{1}{2} [ \omega^2  \sigma^2(r) e^{-\nu }
   + \sigma'^2(r) e^{-\mu } - U ] \; , \\
T_2{}^2  =   T_3{}^3  =p_\bot
=   - \frac{1}{2} [ \omega^2  \sigma^2(r) e^{-\nu }
   - \sigma'^2(r) e^{-\mu } - U ]
\een
where $'=d/dr$. One interesting characteristic of this
system is that the pressure is anisotropic; thus, there are two
equations of state
$p_{r} = \rho - U$ and $p_\bot = \rho - U - \sigma'^2(r) e^{-\mu }$.
The non-vanishing independent components of the Einstein equation are
\ben
\nu' + \mu' & = & \frac{8\pi}{m_{{\rm Pl}}^2} (\rho + p_r) r e^\mu
\; , \label{nula}\\
\mu' & = &  \frac{8\pi}{m_{{\rm Pl}}^2} \rho r e^\mu - \frac {1}{r}
(e^\mu - 1)
\; .
\een
Finally, the scalar field equation is
\be
\sigma'' + \left ( \frac {\nu' - \mu'}{2} + \frac {2}{r} \right )
 \sigma' + e^{\mu - \nu } \omega^2 \sigma
- e^{\mu } \frac{dU}{d\sigma^2} \sigma = 0  \; . \ee To do
numerical computations and order of magnitude estimates, it is
useful to have a new set of dimensionless variables. We adopt here
\be x=mr, \ee for the radial distance, we redefine the radial part
of the boson field as \be \sigma = \sqrt{4\pi} \; \sigma/m_{{\rm
Pl}},\ee and introduce \be \Lambda = \lambda m_{{\rm Pl}}^2/4\pi
m^2,\;\;\; \Omega=\frac {\omega}{m} . \ee In order to obtain
solutions which are regular at the origin, we must impose the
following boundary conditions $\sigma '(0)=0$ and $\mu (0)=0$.
These solutions have two fundamental parameters: the
self-interaction and the central density (represented by the value
of the scalar field at the center of the star). The mass of the
scalar field fixes the scale of the problem. Boundary conditions
representing asymptotic flatness must be applied upon the metric
potentials, these determine --what is actually accomplished via a
numerical shooting method-- the initial value of $\nu=\nu (0)$.
Then, having defined the value of the self interaction, or
alternatively, the form of the soliton potential, the equilibrium
configurations are parameterized by the central value of the boson
field. As this central value increases, so does the mass and
radius of the the star. This happens until a maximum value is
reached in which the star looses its stability and disperses away
(the binding energy being positive). Up to this value of
$\sigma_0$, catastrophe theory can be used to show that these
equilibrium configurations are stable \citep{kus91}. As an example
of boson star configurations, we show in Fig. \ref{conf-1}, the
mass and number of particles (see below) for a $\Lambda=10$ Colpi
et al.'s potential, and in Fig. \ref{conf-2}, the stability
analysis.

When $\Lambda \gg 1$, as in some of the cases we explore in the
next sections, we must follow an alternative adimensionalization
\citep{col86}. For large $\Lambda$, we shift to the following set
of variables, \be \sigma_* = \sigma \Lambda^{1/2},\ee
\be
x_* = x \Lambda^{-1/2},\ee
\be
M_* = M \Lambda^{-1/2}.\ee Here, $M_*$ is defined by
\be
e^\mu= \left( 1-2 \frac{M_*}{x_*} \right)^{-1},\ee which
corresponds to the Schwarzschild mass (see below). Ignoring terms
${\cal O} (\Lambda^{-1})$, the scalar wave equation is solved
algebraically to yield, \be \sigma_*=(\Omega^2 e^{-\nu} -1
)^{1/2},\ee and up to the same accuracy, the field equations are
\be \frac{dM_*}{dx_*} = \frac 14 x_*^2 (3\Omega^2 e^{-\nu} + 1 )
(\Omega^2 e^{-\nu} - 1),\ee
\be
\frac{d\nu}{dx_*} \frac {e^{-\mu}}{x_*} - \frac 1x_*^2 (1-
e^{-\mu} ) = \frac 12 (\Omega^2 e^{-\nu} - 1)^2.\ee The system now
depends only on one free parameter $\Omega^2 e^{-\nu(0)}$.
Numerical solutions show that the maximum mass corresponding to a
stable star is given by $M_{{\rm max}} \sim 0.22 \Lambda^{1/2}
m_{{\rm Pl}}^2/m$.

\subsection{Masses estimates}

The invariance of the Lagrangian density under a global phase
transformation $\psi \rightarrow \psi e^{-i\vartheta }$ of the
complex scalar field gives (via the Noether's theorem) a locally
conserved current $\partial_\mu j^\mu =0$, and a conserved charge
(number of particles). We need to study the number of particles
because it is essential to determine whether the configurations
are stable or not. A necessary requirement towards the stability
of the configurations is a negative binding energy ($BE=M-mN$),
i.e. the star must be energetically more favorable than a group of
unbound particles of equal mass. From the Noether theorem, the
current $j^\mu$ is given by
\be
j^\mu = \frac {i}{2} \sqrt{\mid g\mid }\; g^{\mu \nu }
 [\psi^\ast \partial_\nu \psi -\psi \partial_\nu \psi^\ast ] \; .
\ee and the number of particles is
\be
N := \int j^0  d^3x. \ee For the total gravitational mass of
localized solutions, we may use Tolman's expression \citep{tol34},
or equivalently, the Schwarzschild mass:
\be
M  =  \int (2T_0^{\; 0}-T_\mu^{\; \mu })
         \sqrt{\mid g\mid} \; d^3x \; .
\ee

In Fig. \ref{fitmass}, we show the mass of a boson star with
$\Lambda=0$ as a function of $x$. We have fitted this curve with a
Boltzmann-like function \be M_{{\rm fit}}(x)= A_2 + \frac {A_1 -
A_2} {1 + e^{(x-x_0)/\Delta x} },\ee which is reliable except in
regions very near the center, where $M_{{\rm fit} }(x)$ becomes
slightly negative. The $\chi^2$-parameter of the fitting is around
$10^{-5}$, and values for $A_1, A_2,$ and $\Delta x$ are given in
the figure. It is this formula what we have used in Fig.
\ref{2deTSI} to analytically get the $\sigma_r(p)$ dependence of
the boson star (in the range we use the approximation, the actual
mass and the fitting differ negligibly). Note also what the
fitting is physically telling us: it represents a black hole of
mass $A_1$ plus an inner exponentially decreasing correction. This
is the first time that such a Boltzmann-like fitting is done, and
we think it could be usefully applied in other analytical
computations. Models with different $\Lambda$ and $\sigma$ can be
equally well fitted.

Since boson stars are prevented from gravitational collapse by the
Heisenberg uncertainty principle, we may make some straightforward
mass estimates \citep{mie98}: For a boson to be confined within
the star of  radius $R_0$, the Compton wavelength has to satisfy
$\lambda_\psi= (2\pi\hbar/mc) \leq 2R_0$. In addition, the star
radius must  be of the order of the last stable Kepler orbit
$3R_{\rm S}$ around a black hole of Schwarzschild radius $R_{\rm
S}:= 2GM$. In the case of a mini-boson star of effective radius
$R_0 \cong (\pi/2)^2 R_{\rm S}$ close to its Schwarz\-schild
radius one obtains the estimate
\be
M_{\rm crit} \cong (2/\pi)m_{\rm Pl}^2/m \geq  0.633\, M_{\rm
Pl}^2/m\, . \ee The exact value in the second expression was found
only numerically.

For a mass of $m=30$ GeV, one can estimate the total mass of this
mini--boson star to be $M\simeq 10^{10}$ kg and its radius
$R_0\simeq 10^{-17}$ m, amounting a density 10$^{48}$ times that
of a neutron star. In the case of a boson star ($\lambda \neq 0$),
since $ \vert \psi\vert\sim  m_{\rm Pl}/\sqrt{8\pi}$ inside the
boson star \citep{col86}, the energy density is
\be
\rho \simeq m^2 m_{\rm Pl}^2\left(1 + \Lambda/8\right) \, . \ee
Equivalently, we may think that this corresponds to a  star formed
from  non--interacting bosons with re-scaled mass $m\rightarrow
m/\sqrt{1 +  \Lambda/8} $ and consequently, the maximal mass
scales with the coupling constant $\Lambda$ as,
\be
M_{\rm crit} \simeq {2\over\pi}\sqrt{1 +  \Lambda/8} {m_{\rm Pl}^2
\over m}. \ee There is a large range of values of mass and radius
that can be covered by a boson star, for different values of
$\Lambda$ and $\sigma_0$. For instance, if  $m$ is of the order of
the proton mass and $\lambda \simeq 1$, this is in the range of
the Chandrasekhar limiting mass $M_{\rm Ch}:=M_{\rm
Pl}^3/m^2\simeq 1.5 M_\odot$.

Larger than these estimates is the range of masses that a
non-topological soliton star produces, this is because the power
law dependence on the Planck mass is even higher: $\sim 10^{-2}
(m_{\rm Pl}^4/ m \Phi_0^2)$. These configurations are static and
stable with respect to radial perturbations. It is by no means
true, however, that these are the only stable equilibrium
configurations that one can form with scalar fields. Many
extensions of this formalism can be found. It is even possible to
see that boson stars are a useful setting where to study
gravitation theory in itself \citep{tor97,tor98c,tor98d,whi99}.
Most importantly for our hypothesis is that rotating stable
relativistic boson stars can also be found with masses and radii
comparable in magnitude to their static counterparts
\citep{sch96,yos97}. In astrophysical settings it is usual to
expect some induced rotation of stellar objects, and it is
important that these rotation may not destabilize the structure.
Other interesting generalization is that of electrically charged
boson stars introduced by Jetzer and van der Bij (1989). Although
it is usually assumed that selective accretion will quickly
discharge any astrophysical object, some recent results by Punsly
(1998) suggest this may not always be the case.

From these simple considerations, we can set a difference between
boson and fermion condensations (e.g. neutrino condensation): in
the second case we can have an extended object (in the case of Sgr
A$^*$, its size can be  of about $\sim 30$ld) and it can be quite
diluted. In the case of a boson star, the object is ``strongly''
relativistic, extremely compact, and its size is comparable with
its Schwarzschild radius. We shall discuss the consequences of
this point widely in the next section.

\subsection{Effective potentials}

The motion of test particles can be obtained from the
Euler-Lagrange equations taking into account the conserved
canonical momenta \citep{sha83}. In a general spherically
symmetric potential, the invariant magnitude squared of the four
velocity ($u^2 \equiv g_{\mu\nu} \dot x ^\mu \dot x^\nu$), which
is 1 for particles with non-zero rest mass and 0 for massless
particles, yields to \be \dot r^2= \frac{1}{g_{rr}}\left[
\frac{E_\infty^2}{g_{tt}} -u^2-\frac{l^2}{g_{\phi\phi}}
\right],\ee where  $l$ and $E_\infty$ are constants of motion
given by $-g_{\phi \phi} \sin^2 \theta \; \dot \phi$ and $g_{tt}
\dot t$ (angular momentum and energy at infinity, both per unit
mass) respectively, and a dot stands for derivation with respect
to an affine parameter. This equation can be transformed to
\be
\frac 12 \dot r^2 + V_{ {\rm eff}} = \frac 12 E_\infty^2
e^{-\mu-\nu}, \ee where $V_{{\rm eff}}$ is an effective potential.
This name comes from the fact that in the Schwarzschild solution
$e^{-\mu-\nu}\equiv 1$ and thus, the previous equation can be
understood as a classical trajectory of a particle of energy
$E_\infty^2 /2$ moving in a central potential. This is not so in a
more general spherically symmetric case, like these non-baryonic
stars. In the boson star case, for instance, typical metric
potentials are shown in Fig. \ref{metric}; note that for them
$e^{-\mu-\nu} \neq 1$ We can see, however, that $e^{-\mu-\nu} <
e^{-\mu(0)-\nu(0)} = C$, where $C$ is a constant, and then, usual
classical trajectories can be looked at, in the sense that we may
construct an equation of the form $\dot r^2 /2 + V_{{\rm eff}} <
1/2 E_\infty^2 e^{-\mu(0)-\nu(0)}$, and it will be always
satisfied.

The effective potential that massive or massless particles would
feel must be very different from the black hole case. In the case
of a massless particle, the effective potential is given by
\be
V_{{\rm eff} }=e^{-\mu} \frac{l^2}{2r^2}, \ee while for a massive
test particle it is
\be
V_{{\rm eff}}=\frac 12 e^{-\mu} \left( 1 + \frac{l^2}{r^2}
\right). \ee We show both, boson (for the case of $\Lambda=0$) and
black hole potentials in Figs. \ref{eff-1} and \ref{eff-2}. The
mass of the central object is fixed to be the same in both cases
(see captions) and the curves represent a fourth order Runge-Kutta
numerical integration of the equations we previously derived with
$\Lambda=0$ and $\sigma(0)=0.1$. In the case of massive test
particles we use $l^2/M^2 = 0, 12$ and 15; thus explicitly showing
the change in the behavior of $V_{{\rm eff} }$ for the black hole
case. As $l$ increases, this shape changes from a monotonous
rising curve to one that has a maximum and a minimum before
reaching its asymptotic limit. These extrema disappear for
$l^2/M^2 < 12$. In the case of boson stars, however, if $l \neq 0$
we have a divergence in the center at $r=0$ and only one extremum
-a minimum-, which occurs at rising values of $r/M$ as $l$ grows.
The curve $V_{{\rm eff} }$ for $l=0$ -radially moving objects- is
not divergent, and we can see that the particles may reach the
center of the star with a non-null velocity, given by $1/2\dot r^2
= 1/2 E_\infty^2 e^{-\mu(0)-\nu(0)} - V_{{\rm eff} }(0)
>0$, and will then traverse the star unaffected.

For massless particles, differences are also notorious: a black
hole produces a negative divergence and a boson star a positive
one. Radial motion of massless particles is insensitive to
$V_{{\rm eff}}$, being this equal to zero, as in the Schwarzschild
case. In both cases, for massive and massless particles, outside
the boson star the potential mimics the Schwarzschild one.

\subsection{Particle orbits}

In the case of a black hole, we can use the Newtonian analogy.
Orbits can be of three types: if the energy is bigger than the
effective potential at all points, particles are captured. If the
energy is such that the energy equals the effective potential just
once, particles describe an unbound orbit, and the point of
equality is known as turning point. If there are two such points,
orbits are bound around the black hole. Orbits in which the energy
equals the potential in a minimum of the latter are circular and
stable ($\dot r=0$, $V_{{\rm eff}}^{\prime\prime}<0$).

We can not, because of the different relationship between the
metric potentials we commented above, do the same analysis using
$V_{{\rm eff}}$ for the boson star. We can, however, note that in
most cases the total equation for the derivative of $r$ will be
modified in a trivial way: If we look at the cases where the
effective potential has a divergence at the center (and because
the metric coefficients do not diverge), there is no other
possibility for the particles more than found a turning point.
Orbits can then be bound or unbound depending on the energy, but
we can always find a place where $\dot r =0$ and then it has to
reverse its sign. In the particular cases in which the equality
happens at the minimum of the potential we have again stable
circular orbits. Only in the case where $l=0$ particles can
traverse the scalar star unaffected. For $m \neq 0$ there is still
the possibility of finding a turning point, if the energy is low
enough. However, if the particle is freely falling from infinity,
with $E_\infty = 1$, all energy is purely rest mass, it will
radially traverse the star, as would do a photon.

We conclude that all orbits are not of the capture type. They can
be circular, or unbound, and they all have at least one turning
point. This helps to explain why a non-baryonic object will not
develop a singularity while still being a relativistic object
(comparable effective potentials, equivalently, comparable
relativisticity coefficient $GM(r)/r$.).

\section{Galactic parameters and mass}

One interesting fact, which seems to be not referred before, is
the point that for all these scalar stars, their radius is always
related with the mass in the same way: $R \gtrsim M m_{\rm
Pl}^{-2} $. This is indeed the statement -contrary to what it is
usually assumed- that not all interesting astrophysical ranges of
mass and radius can be modeled with scalar fields. In the scalar
star models, from the given central mass, the radius we obtain for
the star is comparable to that of the horizon, $R \sim m_{\rm
Pl}^{-2} \times 2.61 \times 10^6 M_\odot \sim 3.9 \times
10^{11}$cm.

The question now is for which values of the parameters we can
obtain a scalar object of such a huge mass. For the case of
mini-boson star we need an extremely light boson: \be m[{\rm
GeV}]=1.33 \times 10^{-25} \frac{M(\infty)}{M_{{\rm BH}}}.\ee
Given a central density $\sigma(0)$, $M(\infty)$ stands for the
dimensionless value of the boson star mass as seen by an observer
at infinity. $M_{{\rm BH} }$ is the value of the black hole mass
(in millions of solar masses), obtained by fitting observational
data. Then, we are requiring that the total mass of the boson star
equals that of the black hole. For instance, in Fig. \ref{fiteo},
we have taken Eckart and Genzel's (1997) and Genzel et al.'s
(1996) data, and fit them with a mini-boson star with
$\sigma(0)=0.1$, which yields to a boson mass given by $m[{\rm
GeV}]=2.81 \times 10^{-26}$; the total mass of the star (without
the cluster contribution) is 2.5 $\times 10^6$ M$_\odot$.

In the case of boson stars, and using the critical mass
dependence, $\propto \sqrt \lambda m_{{\rm Pl}}^3/m^2$, the
requirement of a 2.5 million mass star yields to the following
constraint,
\be
m{\rm [GeV]} = 7.9 \times 10^{-4} \left(\frac{ \lambda}{4 \pi}
\right)^{1/4}. \label{bg}\ee It is possible to fulfill the
previous relationship, for instance, with a more heavy boson of
about 1 MeV and $\lambda \sim 1$. A plot of this relation -for
some values of $\lambda$- is shown in Fig. \ref{bos-1}.  Note that
in this case, the value of the dimensionless parameter $\Lambda$
is huge, and special numerical procedures, as explained above,
must be used to obtain solutions. The characteristics of these
solutions have proven to be totally similar to those with $\Lambda
= 0$ which were used in Fig. \ref{fiteo}, just the
adimensionalization differs.

Finally, in the case of a non-topological soliton we obtain the
following constraint,
\be
m{\rm [GeV]} = \frac{7.6 \times 10^{12}}{\Phi_0^2 {\rm [GeV]}^2 }
. \ee For the usually assumed case, in which the order parameter
$\Phi_0$ is of equal value than the boson mass, we need very heavy
bosons of unit mass $m = 1.2 \times 10^{4}$GeV. Other possible
pairs are shown in Fig. \ref{bos-2}.

\subsection{Boson candidates?}

Based only on the constraints imposed by the mass--radius
relationship valid for the scalar stars analyzed, we may conclude
that:
\begin{enumerate}
\item if the boson mass is comparable
to the expected Higgs mass (hundreds of GeV), then the Center of
Galaxy could be a non-topological soliton star;
\item an intermediate mass boson could produce a super-heavy object
in the form of a boson star;
\item for a mini-boson star to be used as
central objects for galaxies it is needed the existence of an
ultra-light boson.
\end{enumerate}
These conclusions should be considered as order of magnitude
estimations. Several reasons force us to make this warning.
Firstly, we are just considering static and uncharged stars, this
is just a model (the simplest), but more complicated ones can
modify the actual constraints. Secondly, we do not know the exact
form of the self-interaction, or in the case of non-topological
stars, the value of $\Phi_0$. For instance, consider the Higgs
mass. In the electro-weak theory a Higgs boson doublet $(\Phi^+,
\Phi^0)$ and its anti-doublet $(\Phi^-, \bar \Phi^0)$ are
necessary ingredients in order to generate masses for the $W^{\pm
}$ and $Z^0$ gauge vector bosons. Calculations of two--loop
electro-weak effects have lead to an indirect determination of the
Higgs mass \citep{gam98}. For a top quark mass of $M_{\rm t}=173.8
\pm 5$ GeV, the Higgs mass is $m_{\rm h}= 104^{+93}_{-49}$ GeV.
However, experimental constraints are weak. Fermilab's tevatron
\citep{han99} has a mass range of $135< m_{\rm h} < 186$ GeV, and
together with LHC at Cern, they could decide if these Higgs
particles (in the given range) exist in nature. Interesting is to
note that as a free  particle, the Higgs boson is unstable with
respect to the decays $h \rightarrow W^+ + W^-$ and $h \rightarrow
Z^0+ Z^0$. However, as remarked by Mielke and Schunck (2000), in a
compact object, and in full analogy with neutron stars --where
there is equilibrium of $\beta $ and inverse $\beta $ decay--
these decay channels are expected to be in equilibrium with the
inverse process $Z^0+ Z^0 \rightarrow h$.

We should also mention the possible dilatons appearing in low
energy unified theories, where the tensor field $g_{\mu\nu}$  of
gravity is accompanied by one or several scalar fields. In string
effective super-gravity \citep{fer94}, for instance, the mass of
the dilaton can be related  to the super-symmetry breaking scale
$m_{\rm susy}$ by $m_\varphi \simeq 10^{-3}(m_{\rm susy}/$
TeV)$^2$ eV. Finally, a scalar with a long history as a dark
matter candidate is the axion, which has an expected light mass
$m_{\sigma} = 7.4\times (10^{7}{\rm  GeV}/f_{\sigma})$ eV
$>10^{-11}$ eV with decay constant $f_{\sigma}$ close to the
inverse Planck time. Goldstone bosons have also inferred mass in
the range of eV and less, $m_g < 0.06 -0.3$ eV \citep{ume98}.

If boson stars really exist, they could be the remnants of
first-order gravitational phase transitions and their mass should
be ruled by the epoch when bosons decoupled from the cosmological
background. The Higgs particle, besides its leading role in
inflationary theories, should be the best and natural candidate as
constituent of a boson condensation if the phase transition
occurred in early epochs. A boson condensation should be
considered as a sort of topological defect relic. In this case, as
we have seen, Sgr A$^*$ could be a soliton star. If soft
phase-transitions took place during cosmological evolution (e.g.
soft inflationary events), the leading particles could have been
intermediate mass bosons and so our supermassive objects should be
genuine boson stars. If the phase transitions are very recent, the
ultra-light bosons could belong to the Goldstone sector giving
rise to mini-boson stars.

We shall not discuss here any further the formation processes of
boson stars. The reader is referred to the recent paper by Mielke
and Schunck (2000) and references therein. It is apparent that for
every possible boson mass in the particle spectrum there is a
boson star model able to fit the galactic center constraints, at
least in order of magnitude.

\section{Accretion and luminosity}

\subsection{Relativistic rotational velocities}

For the static spherically
symmetric metric considered here circular orbit geodesics obey,
\ben v_\varphi^2 & = & \frac{r \nu' e^{\nu }}{2} = e^{\nu }
\frac{e^{\mu}-1}{2} +
    \frac{8\pi}{m_{{\rm Pl}}^2} p_r r^2 \frac{e^{\mu +\nu }}{2}
    \nonumber \\
 & \simeq &
    \frac{M(r)}{r} + \frac{8\pi}{m_{{\rm Pl}}^2} p_r r^2 \frac{e^{\mu +\nu }}{2}
    \; .
\een These curves increase up to a maximum followed by a Keplerian
decrease. Liddle and Schunck (1997) found that the possible
rotation velocities circulating within the gravitational boson
star potential are quite remarkable: {\it their maximum reaches
more than one-third of the velocity of light}. Schunck and Torres
(2000) proved that these high velocities are quite independent of
the particular form of the self-interaction and are usually found
in general models of boson stars. For instance, for
$\Lambda=0,300$ of the Colpi et al. 's (1986) choice, $ U_{\cosh
}= \alpha m^2 \bigl[\cosh(\beta \sqrt{|\psi|^2}) - 1\bigr]$, and $
U_{\exp }= \alpha m^2 \bigl[\exp(\beta^2 |\psi|^2)-1\bigr] $, the
maximal velocities are: 122990 km/s at $x=20.1$ for $\Lambda=300$,
102073 km/s at $x=4.1$ for $\Lambda=0$, 104685 km/s at $x=4.2$ for
$U_{\exp}$, and 102459 km/s at $x=5.9$ for $U_{\cosh }$
\citep{sch00}.

With such high velocities, the matter possesses an impressive
kinetic energy, of about 6\% of the rest mass; i.e. to obtain the
required luminosity we would need that about $10^{-8}$ to
$10^{-7}$ solar masses per year be transformed into radiation.
Note that the required matter-radiation transfer is at least two
orders of magnitude smaller than the accretion rate towards Sgr
A$^*$.

The maximum rotational speed is attained well outside the physical
radius of the star, as can be seen by computing the dimensionless
$x$ value for the star radius (e.g. it happens between $x \sim 5$
and 15 for $\Lambda$ going from 0 to 300). It is interesting to
note also that the dependence of the maximum velocity on $\Lambda$
is not very critical, and the same process can be operative with
mini-boson stars. The rotational velocity is dependent on the
central density, increasing with a higher value of $\sigma_0$. To
obtain large rotational velocities, it is needed that the central
density of the star be highly relativistic, for Newtonian
solutions velocities are low and quite constant over a larger
interval. This is consistent with the density constraint of the
dark object in Sgr A$^*$.

\subsection{The black hole danger}

How can one justify that the accretion onto the central object --a
neutrino ball or a boson star-- will not create a black hole in
its center anyway. Interstellar gas and stars, while spiraling
down towards the center of the object, will begin to collide with
each other, and may glue together at the center, what could be the
seed for a very massive black hole. Even a black hole of small
mass can spiral inwards, and if it remains at the center, that
black hole itself could be the seed. On the other hand, stellar
formation of massive stars would yield, after evolution, to a
black hole. Then, we need to consider {\it whether there is a
mechanism that prevents the formation of a very massive baryonic
object -leading inevitably to a black hole- in the center of the
galaxy.} Below we provide an answer to this question.

The key aspect to consider here is disruption. A star interacting
with a massive object can not be treated as a point mass when it
is close enough to the object such that it becomes vulnerable to
tidal forces. Such effects become important when the pericentre
$r_{{\rm min}}$ is comparable to the tidal radius \citep{ree88},
\be r_t= 5 \times 10^{12} M_6^{1/3} \left( \frac {R_*}{R_\odot}
\right) \left( \frac {M_*}{M_\odot} \right)^{-1/3}\; {\rm cm}.\ee
Here, $R_{*,\odot}$ stands for the radius of the star and the sun
respectively, while $M_{*,\odot}$ for the masses. $M_6$ is the
mass of the central object in millions of solar masses. $r_t$ is
the distance from the center object at which $M/r^3$ equals the
mean internal energy of the passing star. Alternatively, we may
compute the Roche criterion, a comparison between the smoothed out
mass of the disruptor and the internal mean energy of the
disruptee \citep{kro99}. This catastrophe radius is \be R_c \sim
R_g \,2\, \left( \frac{M}{10^8 M_\odot} \right)^{-2/3}
(\rho_\star/\rho_\odot)^{-1/3}, \ee where $R_c$ is given in units
of the event horizon, $R_g$, of the black hole, and the falling
star has density $\rho_* $. Only for black holes with masses
smaller than $10^8 M_\odot$ and for certain low density stars -red
giants- we can expect that the disruption happens outside the
event horizon. This is the reason why supplying the material at
lower densities, with the black hole gravity dominating the
situation, can generate more power. Rees has also given an
estimate of how frequently a star enters  this zone. When star
velocities are isotropic, the frequency with which a solar-like
star pass within a distance $r_{{\rm min}}$ is \be \sim 10^{-4}
M_6^{4/3} \left(\frac{N_*}{10^5 {\rm pc}^{-3}} \right)
\left(\frac{\sigma}{100 {\rm km \;\;s}^{-1}} \right)
\left(\frac{r_{{\rm min}}}{r_t } \right) {\rm yr}^{-1},\ee where
$N_*$ is the star density and $\sigma$ the velocity distribution.
Disruptions are rare events that happens once in about 10000
years.

\subsubsection{Scalar stars}

Because of the similar metric potentials, far from the center of
the non-baryonic star, the accretion mechanism will be the same
than that operative in the Schwarzschild case. If a boson star is
in the center of the galaxy, the characteristics of the tidal
radius and the timescale of disruption occurrence will be similar
to those of a black hole of equal mass. Stars falling inwards will
all be disrupted after they approach a minimum radius. Contrary to
black holes of big masses, which can swallows stars as a whole and
disrupt them behind their event horizons, boson stars will disrupt
all stars --most stars outside and a few inside the Schwarzschild
radius of a black hole of equal mass-- at everyone sight.

In the case of black holes, the most recent simulations
\citep{aya00} show that up to 75\% of the mass that once formed
the disrupted star become unbound. For boson stars, we argue that
once the star is disrupted to test particles (with masses
absolutely negligible to that of the central object), and because
there is no capture orbits, all particles follow unbound
trajectories. In this sense, all material is diverted from the
center and the formation of a black hole is avoided.

It could be worth of interest to perform numerical simulations
changing the central object from a black hole to a boson star, to
see the aftermath and the fate of the debris in an actual
boson-star-generated disruption. This, however, could well not be
an easy task: In black hole simulations, the minimum radius is
maintained still far from the center ($\sim 10$ Schwarzschild
radius), where Newtonian or Post-Newtonian approximations are
valid. To actually see the difference between a black hole and a
boson star it could be necessary to attain inner values of radii,
where the behavior is completely relativistic.

One case could merit further attention: the possible spiraling of
a black hole of stellar size. This case may complicate the
situation since a black hole can not be disrupted. However,
differences in masses are so large that it will behave as a test
particle for the boson central potential, and will be also
diverted from the center. Moreover, being of stellar size, it will
be appreciably influenced by other intruding stars, also making it
to left an static position at the boson star center.

Finally, it is worth of interest to study other possible
observational consequences of boson star disruption. The ejecta
now is 100\% of the star mass, and may possible produce some long
term effects in the surrounding medium. In black hole scenarios,
the bound debris will create a flare as it accretes. Does the same
happen here? While we defer this kind of analysis to another work,
we mention that if boson stars exists, disruption, what was once
thought of as an inevitable concomitant of black holes, could now
happen in non-baryonic environments.

\subsubsection{Neutrino balls}

If the center of the galaxy is a neutrino ball, one also has to
obtain a mechanism that prevents the formation of a very massive
baryonic object. However, we have to expect very crucial
differences with a boson star case, caused by the fact that a
neutrino ball is an extended object and that the gravitational
potential is shallower. The first thing to note is that $r_t$ is
well within the neutrino ball, and then, stars will traverse the
exterior parts of the ball without being disrupted. In doing so,
however, the central mass that they see at the center will be less
than the total mass of the ball, and at a distance $r=r_t$ the
mass enclosed is negligible. Disruption can not proceed and other
mechanism have to devised. We note then that the observation of
disruption processes in the center of a galaxy is then indicative
that a neutrino ball is not there.

When the mass enclosed by the neutrino ball is small enough (say,
${\cal O}(10^{3}) M_\odot$), the accretion disk will be unstable.
This happens about 0.1 -- 1 light years from the center. There,
stars which could actually form at a rate of 1 per hundred
thousand years (the actual number will depend on the mass of the
star) will be probably kicked off by intruding stars
\citep{vio00}. The absence of the disruption mechanism makes this
problem worse, since given an enough amount of time, it is hard to
think in compelling reasons by which gas and stars are expelled
from the center.

In a published paper, Tsiklauri and Viollier (1999) suggested that
matter arriving at the center would be diverted in the form of
non-radiating jets generated by pressure of the inner accretion
disk.
However, it is not clear that gravity attractive forces of the
spiraling objects will be smaller than gas pressure exerted by the
disk. Moreover, even when the baryonic mass acquired by the
central object during the entire age of the universe is at least
two orders of magnitude smaller than the mass of the central
neutrino ball, one should be worried if this yields to a black
hole of that mass, since it may be the seed for a further
collapse, or appreciably influence the dynamics. The mechanism
considered in the previous paragraph seems more adequate than this
for solving this question, if that is finally possible. To us, it
is yet an unclear issue in the neutrino ball scenario.

\section{Discussion: how to differentiate among these models?}

One of the easiest things one may think of is to follow the
trajectory of a particular star. This has been done by Munyaneza
et al. \citep{mun00} in the case of a neutrino ball. The
trajectory of S1, a fast moving star near Sgr A$^*$, offers the
possibility of distinguishing between a black hole and a neutrino
condensate, since Newtonian orbits deviate from each other by
several degrees in a period of some years. However, as soon as the
central object is not so extended, as in the boson star case, this
technique is useless (in every case in which the pericentre is far
than the tidal radius), and other forms of detecting their
possible presence have to be devised.

It has been already noted that X-ray astronomy can probe regions
very close to the Schwarzschild radius. It is only observations in
the X-ray band that can study the inner accretion disk as close to
the center as an event horizon would be. Recent results from the
Japanese-US ASCA mission have revealed a broadened iron line
feature that comes from so close to the event horizon that a
gravitationally redshift is observed. This is 10000 times closer
into the black hole than what can be pictured by HST. In
particular, Iwasawa et al. (1996) claimed that ASCA observations
to Seyfert 1 galaxy MCG-6-30-15 got data from 1.5 gravitational
radii, and conclude that the peculiar line profile suggests that
the line-emitting region is very close to a central spinning
(Kerr) black hole where enormous gravitational effects operate. By
the way, this is stating that a neutrino ball can not be the
center of that galaxy. However, as was already noted
\citep{sch97}, a boson star could well be a possible alternative,
and X-ray could be used to map out in detail the form of the
potential well. The NASA Constellation-X \citep{consx} mission, to
be launched in 2008, is optimized to study the iron K line feature
discovered by ASCA and, if they are there, will determine the
black hole mass and spin for a large number of systems. Still,
Constellation-X will provide an indirect measure of the properties
of the region within a few event horizon radii. A definite answer
in this sense will probably be given by NASA-planned MAXIM mission
\citep{maxim}, a $\mu$-arcsec X-ray imaging mission, that would be
able to take direct X-ray pictures of regions of the size of a
black hole event horizon. Both of these space mission will have
the ability to give us proofs of black hole existence, or to
provide evidence for more strange objects, like boson stars.

Very recently, Falcke et al. (2000) have noted that gravitational
lensing observations of very large baseline interferometry (VLBI)
could give the signature to discriminate among these models.
Falcke et al. assumed that the overall specific intensity observed
at infinity is an integration of the emissivity (taken as
independent of the frequency or falling as $r^{-2}$) times the
path length along geodesics. Defining the apparent boundary of a
black hole as the curve on the sky plane which divides a region
where geodesics intersects the horizon from a region whose
geodesics miss the horizon, they noted that photons on geodesics
located within the apparent boundary that can still escape to the
observer will experience strong gravitational redshift and a
shorter total path length, leading to a smaller integrated
emissivity. On the contrary, photons just outside the apparent
boundary could orbit the black hole near the circular photon
radius several times, adding to the observed intensity. This is
what produces a marked deficit of the observed intensity inside
the apparent boundary, which they refer to as the ``shadow'' of
the black hole. The apparent boundary of the black hole is a
circle of radius $27R_g$ in the Schwarzschild case, which is much
larger than the event horizon due to strong bending of light by
the black hole. This size is enough to consider the imaging of it
as a feasible experiment for the next generation of mm and sub-mm
VLBI. While the observation of this shadow would confirm the
presence of a single relativistic object, a non-detection would be
a major problem for the current black hole paradigm.

Concerning the ideas put forward in this work, Falcke et al.'s
shadow concept is appealing. In the case of a boson star, we might
expect some diminishing of the intensity right in the center, this
would be provided by the effect on relativistic orbits, however,
this will not be as pronounced as if a black hole is present: for
that case, many photons are really gone through the horizon and
this deficit also shows up in the middle. If a boson star is
there, some photons will traverse it radially, and the center
region will not be as dark as in the black hole case. A careful
analysis of Falcke et al.'s shadow behavior replacing the central
black hole with a boson star model would be necessary to get any
further detail, and eventually an observable prediction.

We also mention that the project ARISE (Advanced Radio
Interferometry between Space and Earth) is going to use the
technique of Space VLBI to increase our understanding of black
holes and their environments. The mission, to be launched in 2008,
will be based on a 25-meter inflatable space radio telescope
working between 8 and 86 GHz \citep{ulv99}. It will study
gravitational lenses at resolutions of tens of $\mu$arcsecs,
yielding information on the possible existence (and signatures) of
compact objects with masses between $10^3 M_{\odot}$ and $10^6
M_{\odot}$.

Another possible technique for detecting boson stars from other
relativistic objects could be gravitational wave measurements
\citep{rya97}. If a particle with stellar mass is observed to
spiral into a spinning object with a much larger mass and a radius
comparable to its Schwarzschild length, from the emitted
gravitational waves, one could in principle obtain the lowest
multipole moments. The black hole no-hair theorem says that all
moments are determined by its lowest two, the mass and angular
momentum (assuming the charge equal to zero). Should this not be
so, the central object would not be a black hole, and as far as we
know, the only remaining viable candidate would be a boson star.
In ten years time, perhaps, a combination of gravitational wave
measurements, better determination of stellar motions, and mm and
sub-mm VLBI techniques could give us a definite picture of the
single object at the center of our Milky Way.

From a theoretical point of view, developments on gravitational
lensing theory in very strong field regimes will be of extreme
importance since, for objects like Sgr $A^*$, the standard weak
field theory does not hold and new effects have to be expected
\citep{vir00}.

\section{Conclusions}

We have shown that boson stars provide the basic necessary
ingredients to fit dynamical data and observed luminosity of the
center of the Galaxy. They constitute viable alternative
candidates for the central supermassive object, producing a
theoretical curve for the projected stellar velocity dispersion
consistent with Keplerian motion, relativistic rotational
velocities, and having an extremely small size.

Other singularity-free models were as well considered, as
non-baryonic fermion stars (e.g. neutrino, or gravitino
condensations). In this case, the object is sustained by its Fermi
energy, while in the boson star case, it is the Heisenberg's
Uncertainty Principle which prevents the system from collapsing to
a singularity. Due to this fact, boson stars are genuine
relativistic objects where a strong gravitational field regime
holds.

This difference in the relativistic status of both objects is not
trivial. While fermion neutrino balls are extended objects, boson
stars mimic a black hole. Disruption processes can not happen in a
fermion condensation, and it has the unpleasant consequence of not
providing with a straightforward mechanism by which stars could be
diverted from the center, and through which finally avoid the
formation of a massive black hole inside the condensate.

The formation of boson stars, neutrino balls, and black holes can
all be competitive processes. Then, it might well be that even if
we discover that a black hole is in the center of the galaxy,
others galaxies could harbor non-baryonic centers. In the case of
boson stars, only after the discovery of the boson mass spectrum
we shall be in position to determine a priori which galaxies could
be modeled by such a center. Observations of galactic centers
could then suggest the existence of boson scalars much before than
their discovery in particle physicists labs.

\acknowledgments

We acknowledge insightful comments by H. Falcke, F. Schunck, D.
Tsiklauri, R. Viollier, and A. Whinnett. Also, we acknowledge
informal talks with D. Bennet, R. di Stefano, C. Kochanek, and A.
Zaharov at the Moriond 2000 meeting. D.F.T. was supported by
CONICET as well as by funds granted by Fundaci\'{o}n Antorchas,
and particularly thanks G. E. Romero for his encouragement and
criticism. S. C. and G. L.'s research was supported by MURST fund
(40\%) and art. 65 D.P.R. 382/80 (60\%). G.L. further thanks UE
(P.O.M. 1994/1999).

\appendix
\section{Appendix}

For reader's convenience, we quote here the dimensional conversion
for the radius and the mass of a boson star. Using the value of 1
GeV in cm$^{-1}$, and taking into account the dimensionless
parameter $x=mr$, we get \be r[{\rm pc}]= \frac {x}{m[{\rm GeV}]}
6.38 \times 10^{-33}.\ee For the mass, recalling that $M=M(x)
m_{\rm Pl}^2/m$, we get \be M[10^6 M_\odot]=\frac{M(x)}{m[{\rm
GeV}]} 1.33 \times 10^{-25}.\ee In the case where $\Lambda \gg 1$,
both right hand sides of the previous formulae get multiplied by
$\Lambda^{1/2}$.

\clearpage

\figcaption[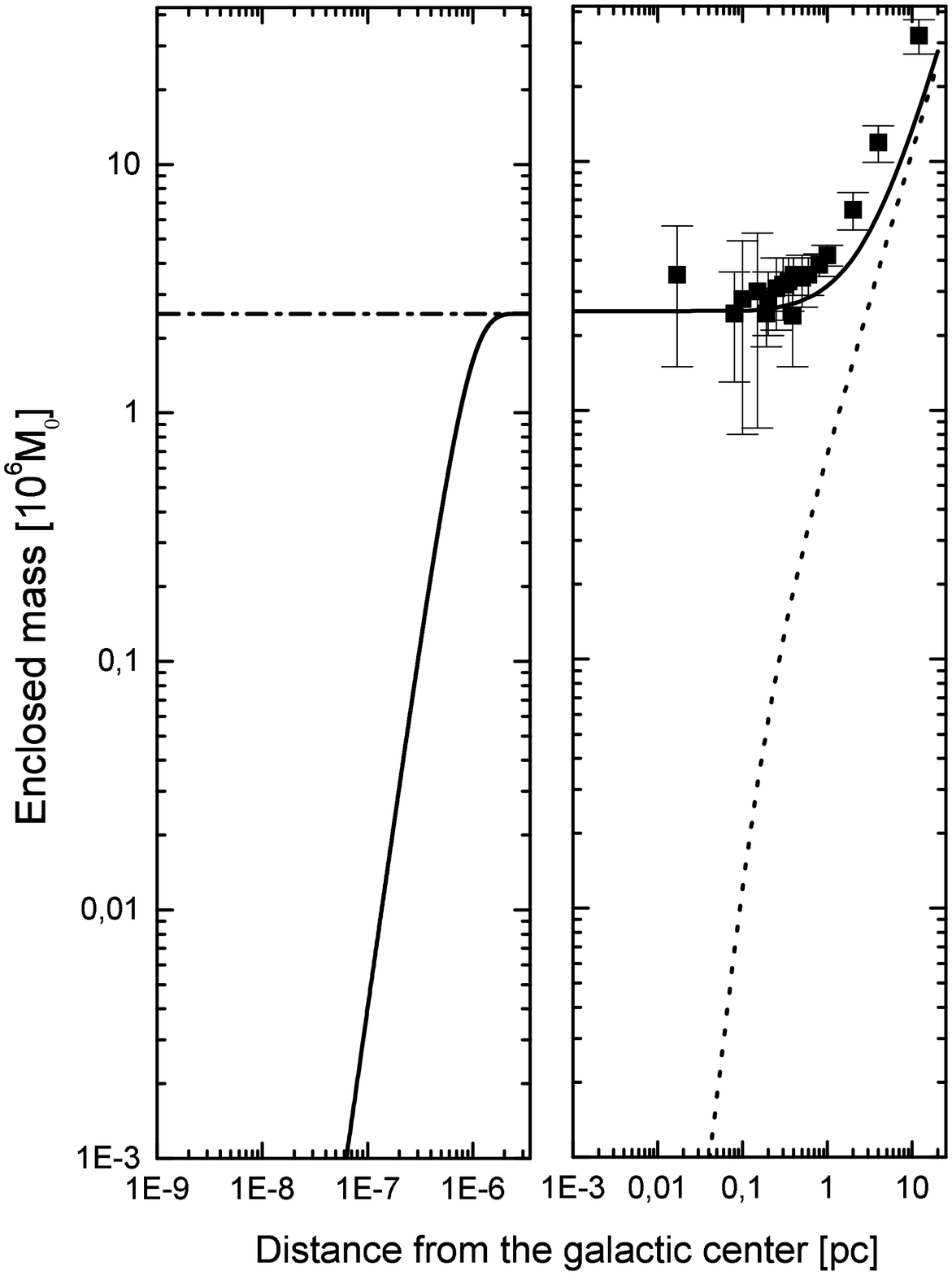] {Enclosed mass in the center of the galaxy
together with observational data points. See discussion in the
main text. \label{fiteo}}

\figcaption[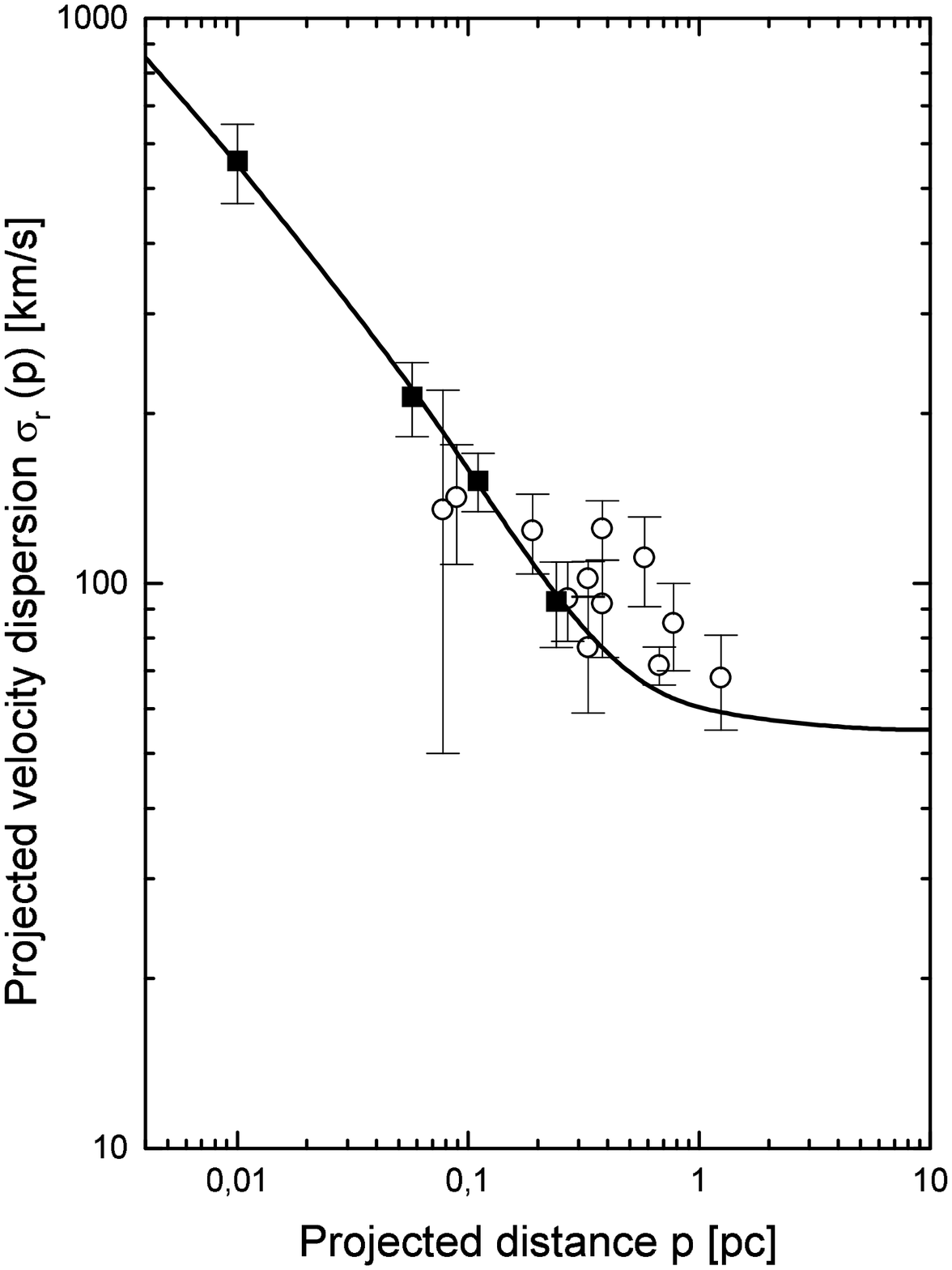] {Projected velocity dispersions:
observational data and fit using a boson star model. See
discussion in the main text. \label{2deTSI}}

\figcaption[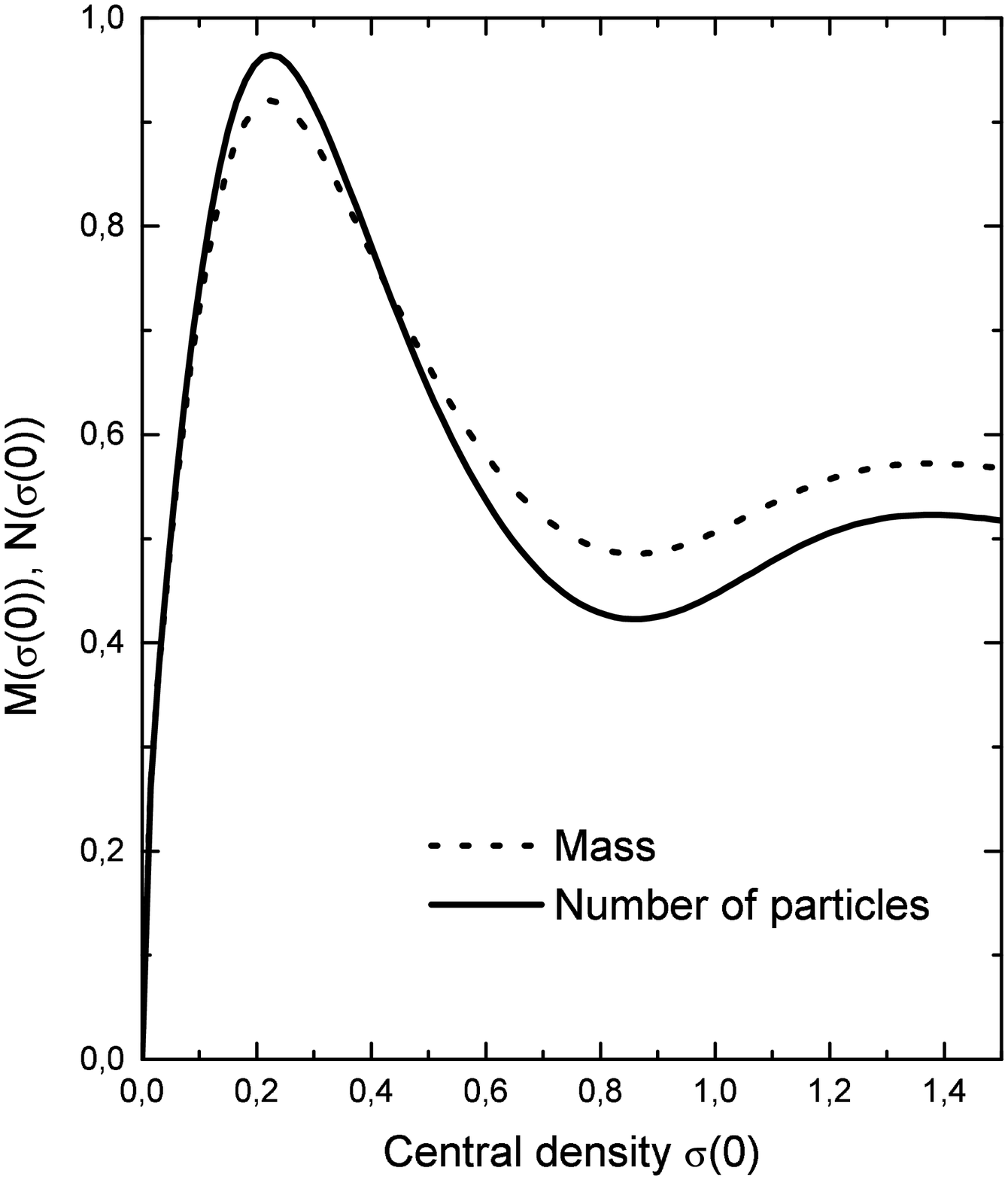]{Mass and number of particles, in
dimensionless units, for a $\Lambda=10$ boson star (Colpi et al.
(1986)) potential. \label{conf-1}}

\figcaption[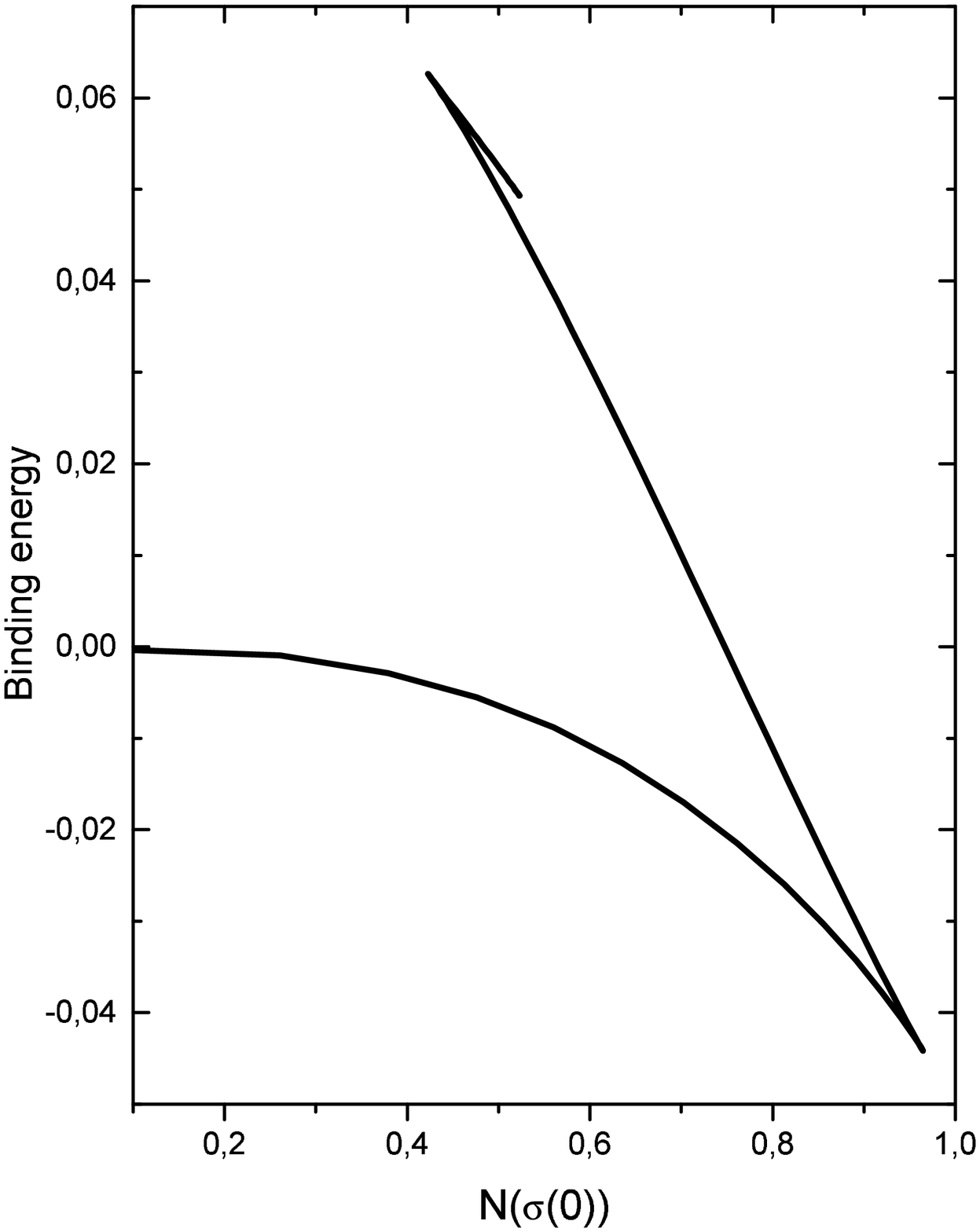]{Stability analysis for $\Lambda=10$
configurations. Catastrophe theory ensures us that the first
branch, which includes the (0,0) point, is the only stable
one.\label{conf-2}}

\figcaption[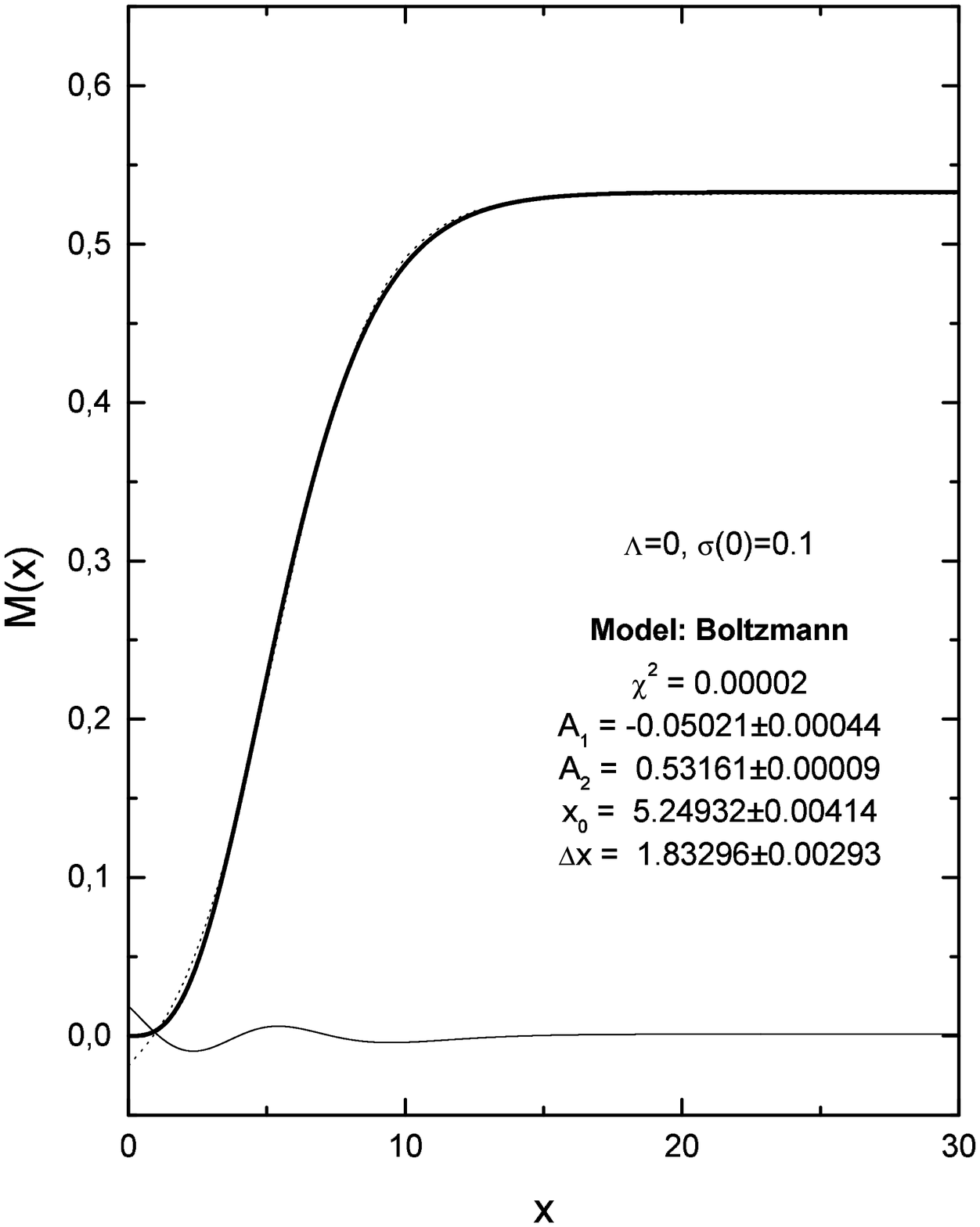]{Boson star mass as a function of $x$, for
a $\Lambda=0$, $\sigma(0)=0.1$ model. We show a Boltzmann-like
fitting (dotted curve, but almost everywhere on the solid curve)
and its residue (solid lower line). \label{fitmass}}

\figcaption[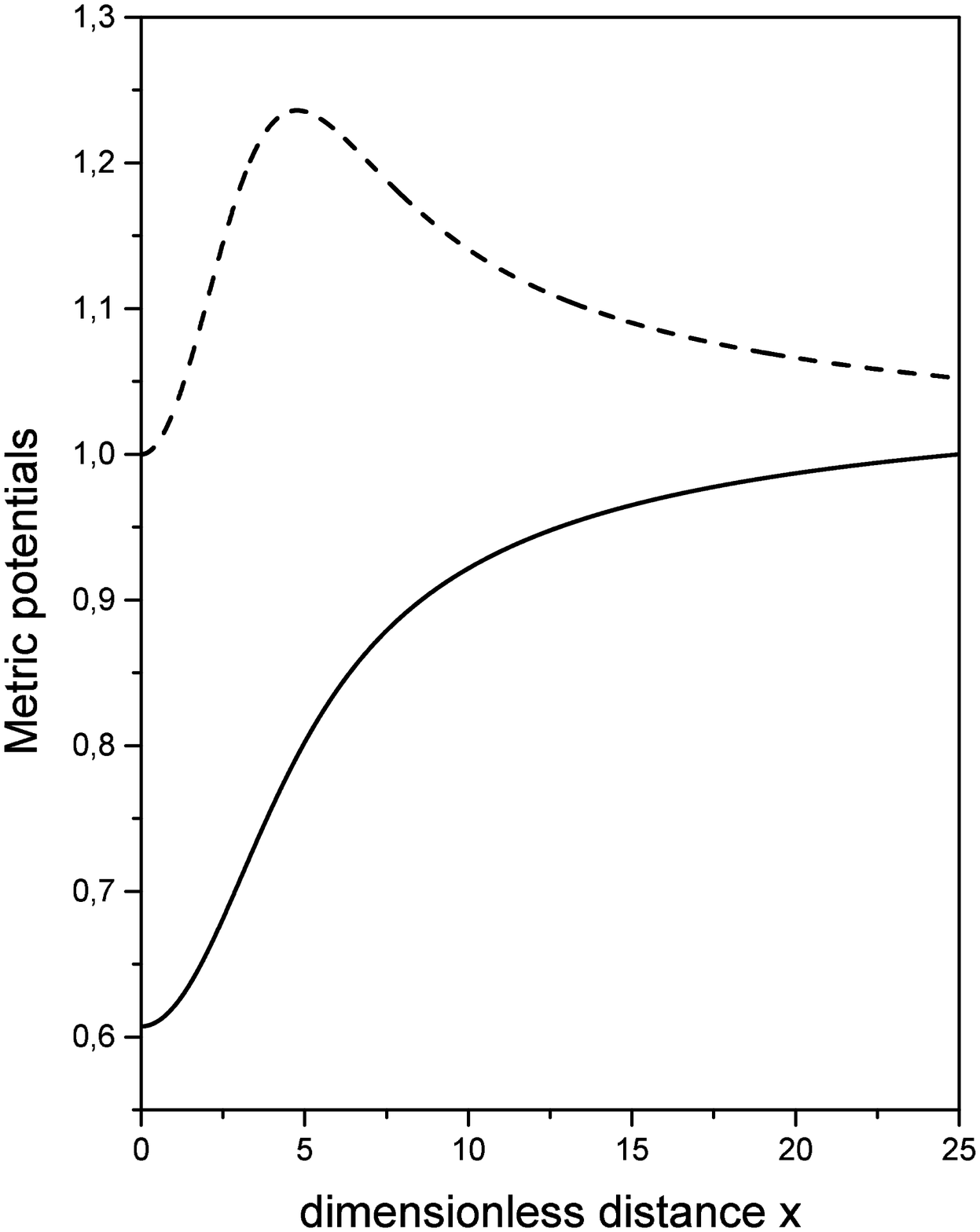]{Boson star metric potentials
$g_{rr}=e^\mu$ (dashed) and $g_{tt}=e^\nu$ (solid). Boson star
parameters are $\Lambda=0$, $\sigma(0)=0.1$. \label{metric}}

\figcaption[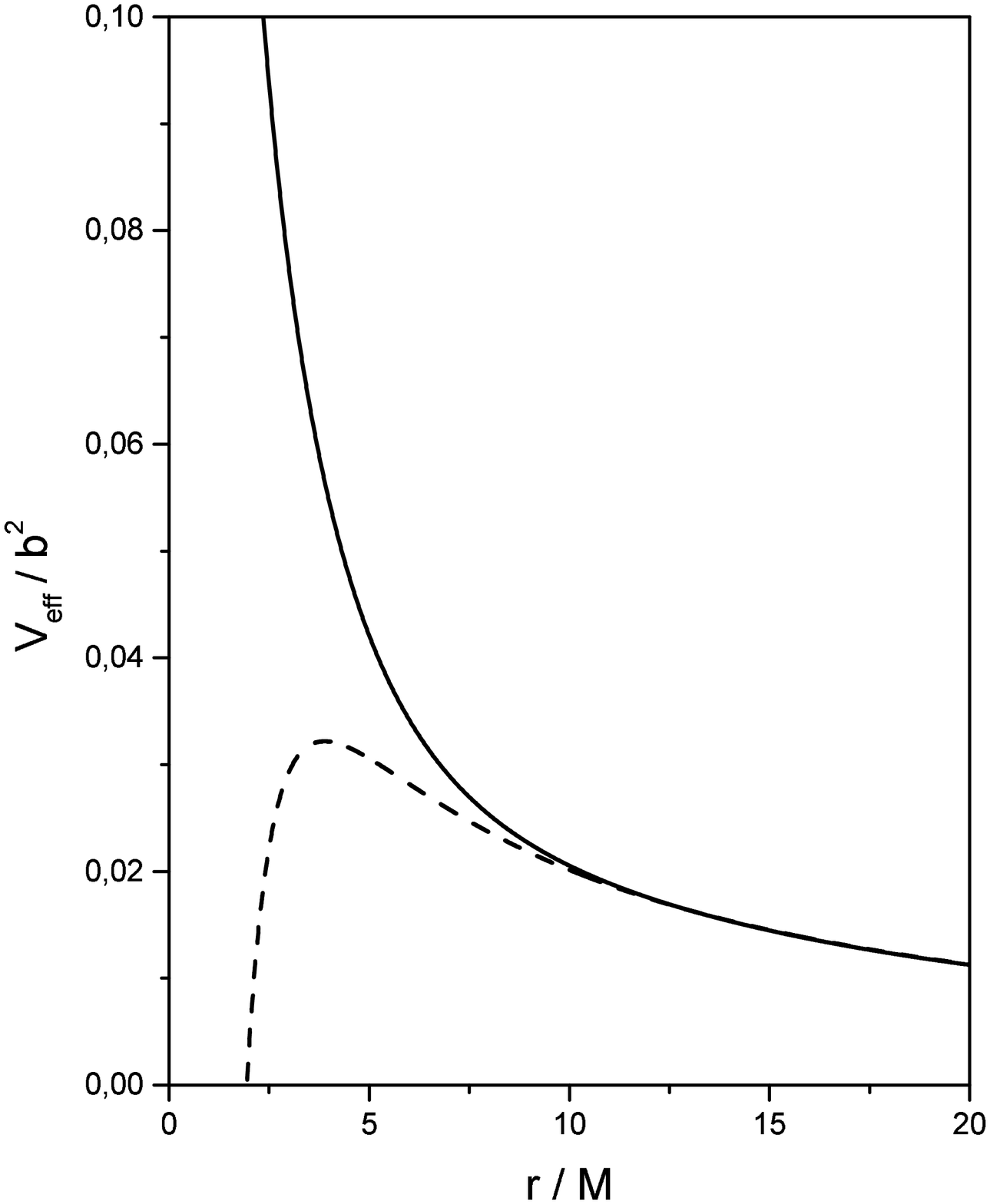]{Boson (solid) and Schwarzschild (dashed)
effective potentials for massless particles, $b=l^2/E_\infty^2$.
The mass for both, the black hole and the boson star, was taken as
$M=0.62089 m_{\rm Pl}^2 / m$. The maximum in the black hole case
happens, independently of $l$, for $r/M=3$. \label{eff-1}}


\figcaption[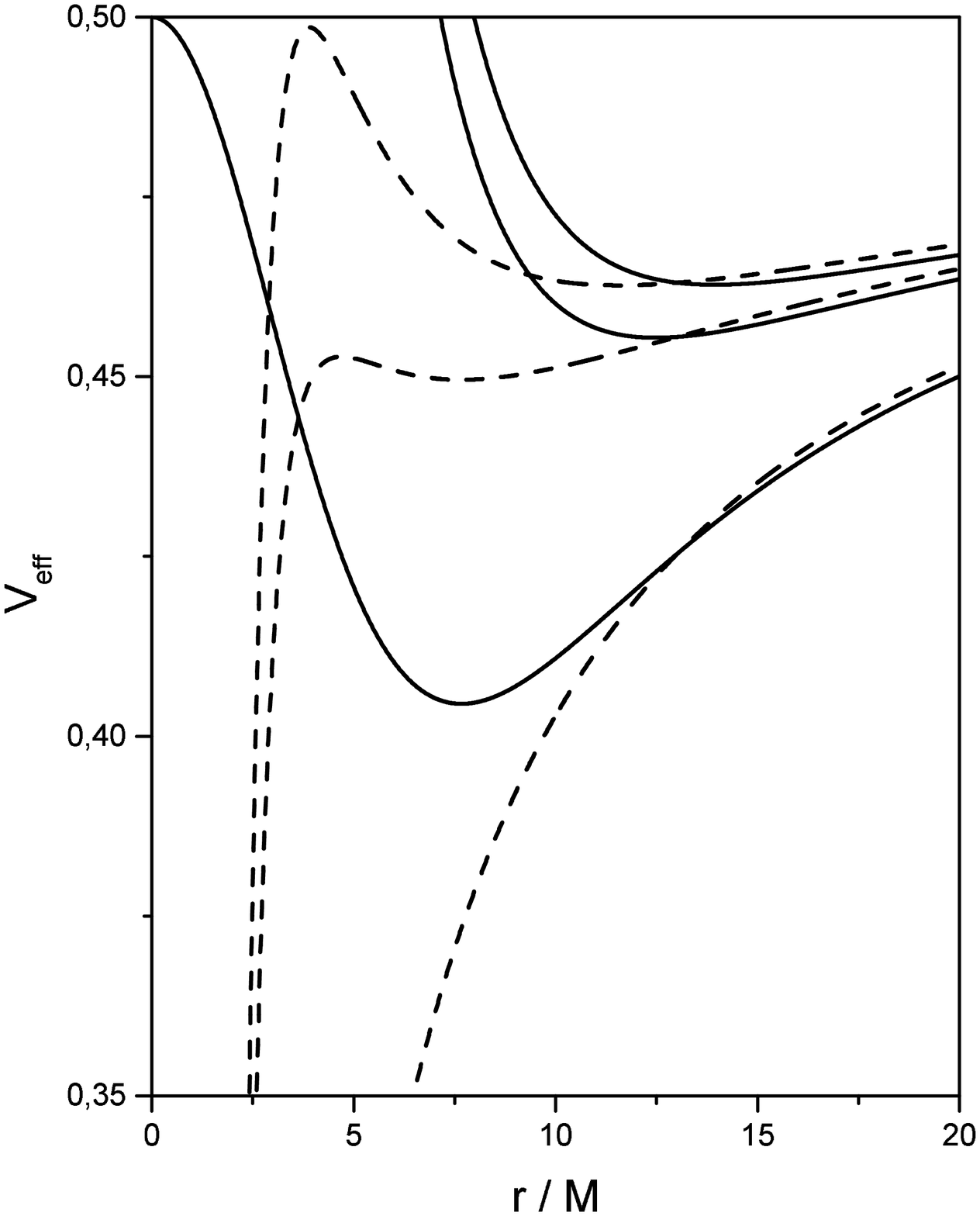]{Boson (solid) and Schwarzschild (dashed)
effective potentials in the case of massive particles. The mass of
the central object is as in the previous figure. The three curves
correspond to $l^2/M^2=0, 12$ and 15 (from bottom to top).
\label{eff-2}}

\figcaption[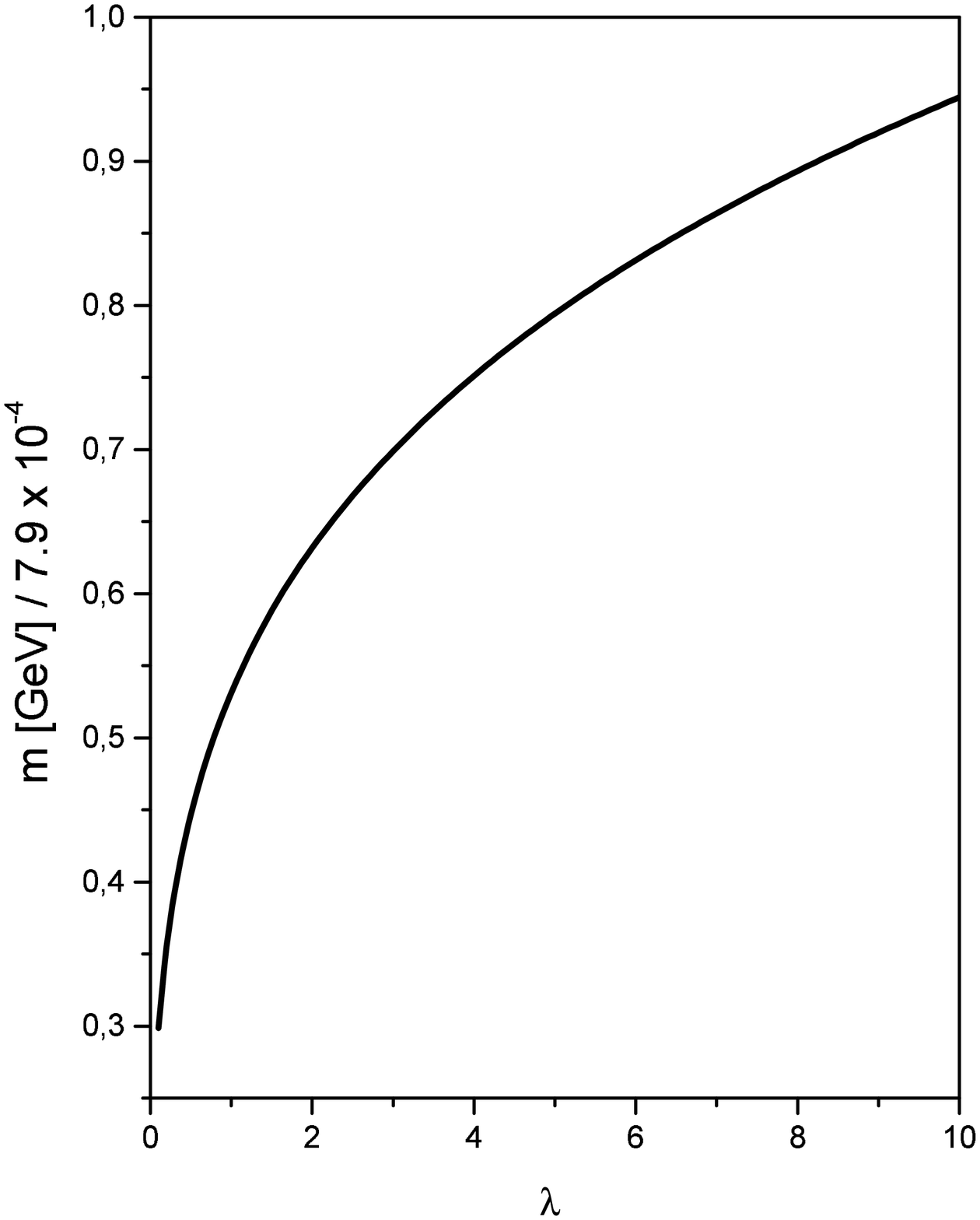]{Constraint in the boson star fundamental
parameters which gives rise to an object of two million solar
masses within approximately ten solar radius, consistent with the
mass of the central object in our galaxy. \label{bos-1}}

\figcaption[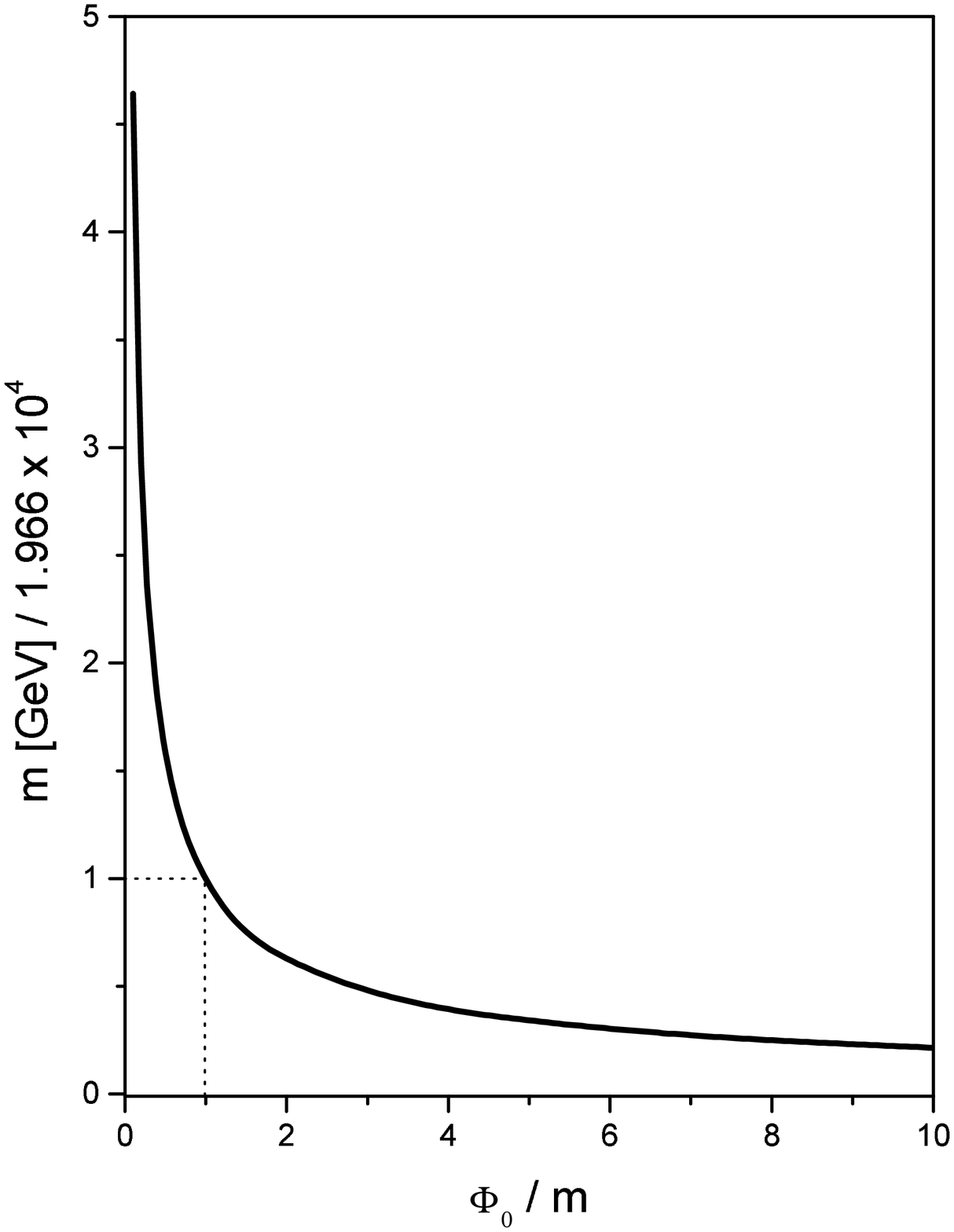]{Constraint in the non-topological soliton
fundamental parameters which gives rise to an object consistent
with the mass of the central object in our galaxy. It is specially
marked the usual case in which $\Phi = m$. \label{bos-2}}

\end{document}